\documentclass[twocolumn]{aastex631}

\newcommand\cd{d$^{-1}$\,}
\newcommand\cms{cm\,s$^{-2}$\,}
\usepackage{multirow}
\usepackage{amsmath}
\usepackage{mathrsfs}
\usepackage[T1]{fontenc}

\accepted{September 19, 2023}

\begin{document}

\title{The GW Vir instability strip in the light of new observations of PG 1159 stars. \\ Discovery of pulsations in the central star of Abell 72 and variability of RX J0122.9$-$7521}

\correspondingauthor{Paulina Sowicka, Gerald Handler}
\email{paula@camk.edu.pl, gerald@camk.edu.pl}

\author[0000-0002-6605-0268]{Paulina Sowicka}
\affiliation{Nicolaus Copernicus Astronomical Center, Polish Academy of Sciences, ul. Bartycka 18, PL-00-716, Warszawa, Poland}

\author[0000-0001-7756-1568]{Gerald Handler}
\affiliation{Nicolaus Copernicus Astronomical Center, Polish Academy of Sciences, ul. Bartycka 18, PL-00-716, Warszawa, Poland}

\author[0000-0003-3947-5946]{David Jones}
\affiliation{Instituto de Astrof\'isica de Canarias, E-38205 La Laguna, Tenerife, Spain}
\affiliation{Departamento de Astrof\'isica, Universidad de La Laguna, E-38206 La Laguna, Tenerife, Spain}
\affiliation{Nordic Optical Telescope, Rambla Jos\'e Ana Fern\'andez P\'erez 7, 38711, Bre\~na Baja, Spain}

\author{John A. R. Caldwell}
\affiliation{McDonald Observatory, 82 Mt. Locke Road, McDonald Observatory, TX 79734, USA}

\author{Francois van Wyk}
\affiliation{South African Astronomical Observatory, PO Box 9, Observatory, 7935 Cape, South Africa}

\author[0000-0002-3304-5200]{Ernst Paunzen}
\affiliation{Department of Theoretical Physics and Astrophysics, Faculty of Science, Masaryk University, Kotlářská 2, Brno, Czech Republic}

\author[0000-0003-1034-1557]{Karolina B\k{a}kowska}
\affiliation{Institute of Astronomy, Faculty of Physics, Astronomy and Informatics, Nicolaus Copernicus University, ul. Grudziądzka 5, 87-100 Toruń, Poland}

\author[0000-0002-3084-084X]{Luis Peralta de Arriba}
\affiliation{Centro de Astrobiolog\'{\i}a (CAB), CSIC-INTA, Camino Bajo del Castillo s/n,
28692 Villanueva de la Ca\~nada, Madrid, Spain}
\affiliation{Isaac Newton Group of Telescopes, E-38700 Santa Cruz de La Palma, La Palma, Spain}

\author[0000-0002-0887-1009]{Lucía Suárez-Andrés}
\affiliation{Isaac Newton Group of Telescopes, E-38700 Santa Cruz de La Palma, La Palma, Spain}

\author[0000-0002-6428-2276]{Klaus Werner}
\affiliation{Institut f\"{u}r Astronomie und Astrophysik, Kepler Center for Astro and Particle Physics, Eberhard Karls Universit\"{a}t, Sand 1, D-72076 T\"{u}bingen, Germany}

\author[0000-0003-0751-3231]{Marie Karjalainen}
\affiliation{Astronomical Institute, Czech Academy of Sciences, Fričova 298, 25165, Ondřejov, Czech Republic}

\author[0000-0003-2002-896X]{Daniel L. Holdsworth}
\affiliation{Jeremiah Horrocks Institute, University of Central Lancashire, Preston, PR1 2HE, UK}


\begin{abstract}

We present the results of new time series photometric observations of 29 pre-white dwarf stars of PG~1159 spectral type, carried out in the years $2014 - 2022$. For the majority of stars, a median noise level in Fourier amplitude spectra of $0.5 - 1.0$ mmag was achieved. This allowed the detection of pulsations in the central star of planetary nebula Abell 72, consistent with g-modes excited in GW Vir stars, and variability in RX J0122.9$-$7521 that could be due to pulsations, binarity or rotation. For the remaining stars from the sample that were not observed to vary, we placed upper limits for variability. After combination with literature data, our results place the fraction of pulsating PG~1159 stars within the GW Vir instability strip at 36\%. An updated list of all known PG~1159 stars is provided, containing astrometric measurements from the recent \textit{Gaia} DR3 data, as well as information on physical parameters, variability, and nitrogen content. Those data are used to calculate luminosities for all PG~1159 stars to place the whole sample on the theoretical Hertzsprung-Russell diagram for the first time in that way. The pulsating stars are discussed as a group, and arguments are given that the traditional separation of GW Vir pulsators in ``DOV'' and ``PNNV'' stars is misleading and should not be used.
\end{abstract}

\keywords{PG 1159 stars (1216), Pulsating variable stars (1307), Stellar pulsations (1625), Non-radial pulsations (1117), Stellar evolution (1599), CCD photometry(208), Hertzsprung Russell diagram(725), Post-asymptotic giant branch(1287)}

\section{Introduction} \label{sec:intro}

Pre-white dwarf stars of PG 1159 spectral type (named after the prototype, PG~1159$-$035, \citealt{1979wdvd.coll..118G}) are important to study in the context of stellar evolution, as they are supposed main progenitors of H-deficient white dwarfs (WDs).
PG~1159 stars populate the GW Vir instability strip, together with central stars of planetary nebulae with C-rich Wolf-Rayet spectra ([WC]-types, exhibiting He, C, and O lines in emission, \citealt{1998MNRAS.296..367C}), and [WC]-PG1159 stars, so-called transition objects \citep{1993AcA....43..329L, 2015ApJ...799...67T}. PG~1159 stars exhibit a broad absorption ``trough'' made by \ion{He}{2} at $4686\,$\AA{} and adjacent \ion{C}{4} lines (see, e.g., Fig.~2 from \citealt{2014AA...569A..99W}), and typically have He-, C- and O-rich atmospheres, but notable variations in He, C, and O abundances were found from star to star (e.g., \citealt{1998AA...334..618D}, \citealt{2001ApSS.275...27W}). Other groups of (pre-) white dwarf stars also show \ion{He}{2} and \ion{C}{4} lines -- while the O(He) stars show significantly less carbon than PG~1159 stars (up to 3\% in their atmospheres, \citealt{2014AA...566A.116R}), the limit to distinguish between PG~1159 stars and DO white dwarfs is model-dependent -- \citet{2014AA...564A..53W} adopted C/He up to 9\% (by mass) for DO stars.

Their formation history involves either a single star evolution scenario -- a ``born-again'' episode (a Very Late Thermal Pulse -- VLTP, or a Late Thermal Pulse -- LTP; PG 1159-hybrid stars experience an AGB Final Thermal Pulse -- AFTP), or binary evolution -- binary white dwarf merger \citep{2022MNRAS.511L..66W,2022MNRAS.511L..60M}. 
Only some stars within the GW Vir instability strip show pulsations, a striking difference from the other two classical white dwarf instability strips (DAV and DBV), which are believed to be pure (e.g., see \citealt{2008PASP..120.1043F}). The GW Vir pulsations are due to nonradial g-modes, where the main restoring force is gravity (buoyancy), driven by the $\kappa-\gamma$ mechanism associated with the partial ionization of the K-shell electrons of carbon and/or oxygen in the envelope. The pulsations typically are of short period (between 300 s and about 6000 s) and low amplitude (typically 1 mmag -- 0.15 mag, \citealt{2019AARv..27....7C}).   
 
A current hypothesis, based on combined photometric and spectroscopic observations, states that there is a clear separation within PG 1159 stars: all N-rich (about 1\% atmospheric N/He abundance) PG 1159 stars are pulsators, while all N-poor ones (below about 0.01\% N/He) do not pulsate \citep{1998AA...334..618D,2021ApJ...918L...1S}. Since N is a tracer of the evolutionary history, an important conclusion follows: the pulsating and non-pulsating PG 1159 stars have different evolutionary histories, and it seems necessary that a star undergoes a VLTP in order to develop pulsations. 
Recently, considerable progress has been made in the study of PG~1159 stars' atmospheric structure, composition, and evolution through optical and ultraviolet spectroscopy and advancement in non-LTE model atmospheres, as well as in probing their interiors through asteroseismology with space-based observations (e.g., the \textit{TESS} mission \citep{2015JATIS...1a4003R} observed several already known GW Vir stars, \citealt{2021AA...645A.117C}).
In the light of these findings, it is important to further test this hypothesis on a larger sample of PG 1159 stars, by obtaining high-quality, high-speed photometric observations aimed at detecting low-amplitude pulsations if present, as well as spectroscopic observations capable of detecting the nitrogen lines.

The number of known PG 1159 stars has increased in recent years, both due to the detection of pulsations typical for GW Vir stars in new photometric surveys (e.g., \textit{TESS}, \citet{2021AA...655A..27U, 2022MNRAS.513.2285U}, confirmed by spectroscopy), and classification of targets of spectroscopic surveys (e.g., the most recent discoveries with HET, \citealt{2023MNRAS.521..668B}). Currently, 67 PG 1159 stars are known\footnote{Based on a list from \cite{2006PASP..118..183W}, updated by us.}, including hybrid-PG 1159 stars (whose atmospheres have traces of hydrogen). While these stars lay within the GW Vir instability strip, some of them were either never checked for (or reported) variability, or the quality of previous observations was not sufficient to detect low amplitude pulsations. They also could have been observed when beating between closely spaced modes was destructive and pushed the observed amplitudes below the detection threshold. Moreover, some of these objects have temporally highly variable pulsation spectra \citep{1996AJ....111.2332C}. Therefore, it is worth re-observing those stars in different observing cycles to look for photometric variability. To date, there was no extensive and systematic photometric survey for variability among those stars since the works of \citet{1987ApJ...323..271G}, \citet{1996AJ....111.2332C}, and \citet{2006AA...454..527G}. 

The aim of the work presented in this paper is to obtain new photometric observations of a selected sample of PG 1159 stars to find new pulsators (or candidates) and put limits on non-variability. We also provide the most up-to-date list of PG 1159 stars and their properties from the \textit{Gaia} mission and follow-up works. Finally, we place the PG~1159 stars on the theoretical Hertzsprung-Russell (HR) diagram ($\log{}L_{\star} / L_{\odot}-\log{}T_{\mathrm{eff}}$) and discuss the implications of our findings.

\newpage

\section{Photometric observations}
\label{sec:obs}

We selected a sample of PG 1159 stars for a survey of variability carried out in the years $2014 - 2022$ with a network of telescopes, covering both hemispheres. The selection was based on only one criterion: a given star was included in our target list if it was never observed photometrically with time resolution sufficient for the detection of GW Vir pulsations, or was classified as non-variable, but either the reported detection limits could have been improved by new observations or were not provided by the previous authors. The top panel in Figure~\ref{fig:hist} shows the brightness distribution of observed stars. 
The observing plan assumed the acquisition of observing blocks lasting at least one hour per target. The following telescopes and instruments were used for observations:
\begin{figure}
	\includegraphics[width=\linewidth]{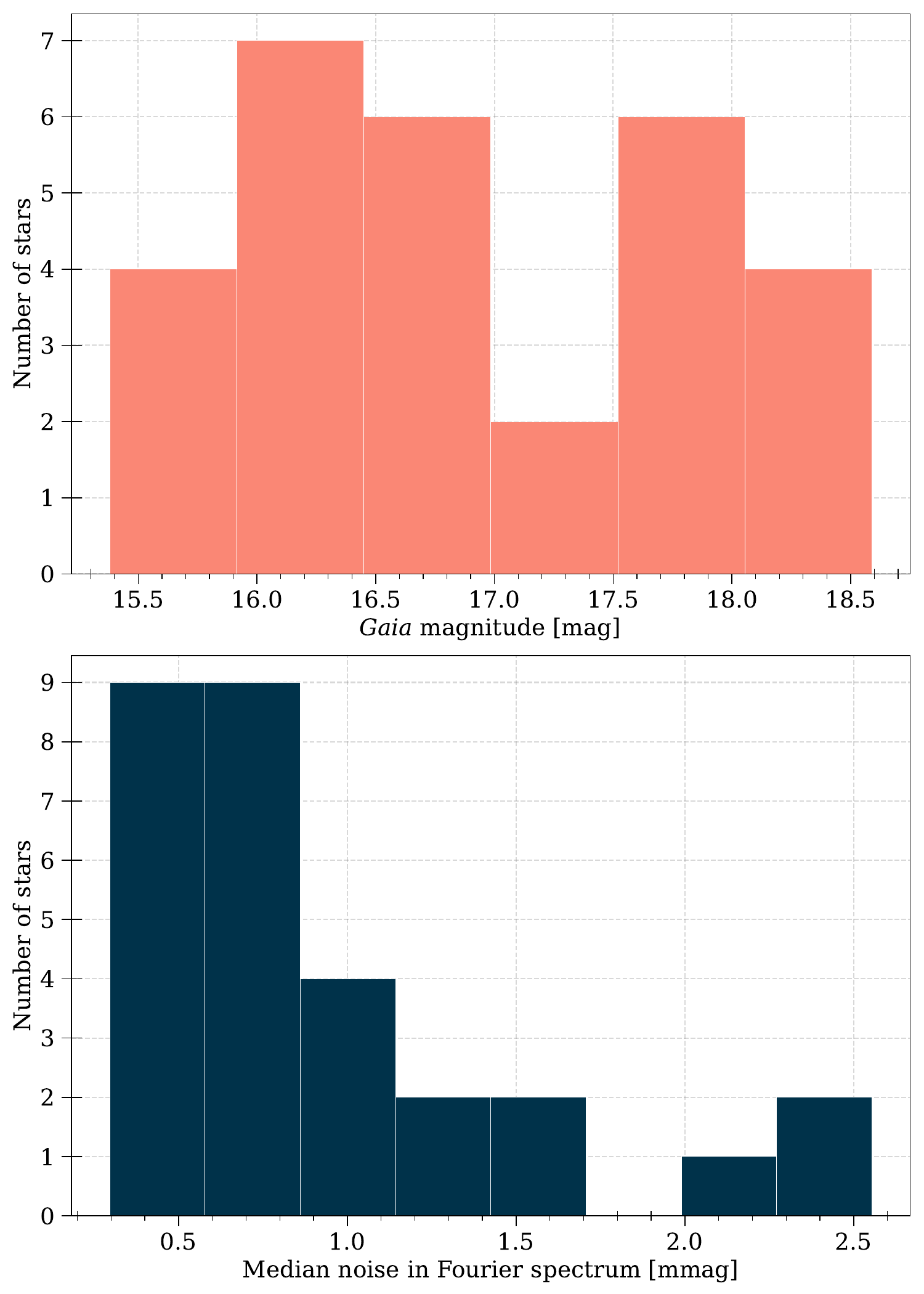}
    \caption{\textit{Top:} \textit{Gaia} magnitude distribution of the observed sample of 29 PG~1159 stars. \textit{Bottom:} Distribution of the median noise level achieved in the survey.}
    \label{fig:hist}
\end{figure}
\begin{itemize}
    \item \textbf{DFOSC at the 1.54-m Danish Telescope at ESO (DK):} \\
    The 1.54-m Danish Telescope located at La Silla Observatory was equipped with the Danish Faint Object Spectrograph and Camera (DFOSC; \citealt{1995Msngr..79...12A}). DFOSC uses a 2K$\times$2K thinned Loral CCD chip with a Field of View (FOV) of 13.7$\times$13.7 arcmin. No filter was used. Six stars were observed with this telescope. %

    \item \textbf{OSIRIS at 10.4-m Gran Telescopio Canarias (GTC):} \\
    The 10.4-m Gran Telescopio Canarias (GTC) is located at the  Observatorio del Roque de los Muchachos (ORM, La Palma) and is equipped with Optical System for Imaging and low-Intermediate-Resolution Integrated Spectroscopy (OSIRIS; \citealt{1998Ap&SS.263..369C}). OSIRIS consists of a mosaic of two CCDs of 2048$\times$4096 pixels each and has an unvignetted FOV of 7.8$\times$7.8~arcmin. Either no filter or a Sloan~r' filter was used. We used 2$\times$2 binning and a standard readout time of about 23~seconds. Eleven stars were observed with this telescope.

    \item \textbf{WFC at 2.54-m Isaac Newton Telescope (INT):} \\
    The 2.54-m Isaac Newton Telescope (INT) is located at the Observatorio del Roque de los Muchachos (ORM, La Palma) and is equipped with the Wide Field Camera (WFC; \citealt{2001INGN....4....7W}), an optical mosaic camera mounted in the prime focus. WFC consists of four thinned EEV 2k$\times$4k CCDs. Because the readout time of the whole CCD mosaic is rather long, we used it in windowing mode -- for a FOV of 5$\times$5 arcmin (910$\times$910 pixels) the readout time was 6~seconds in the slow (less noisy) mode. No binning was used. We used a Harris V filter. Five stars were observed with this telescope.

    \item \textbf{ProEM at the 2.1-m Otto Struve Telescope (MD):} \\
    The 2.1-m Otto Struve Telescope is located at McDonald Observatory, and is equipped with ProEM, which is a frame-transfer CCD detector with optional electron-multiplication with high frame-rate, optimized for high-speed time-series photometry (providing effectively zero readout time). The CCD has 1024$\times$1024 pixels and a FOV of 1.6$\times$1.6 arcmin. We used 4$\times$4 binning for an effective plate scale of 0.36~arcsec pixel$^{-1}$. We used a BG40 filter. Nine stars were observed with this telescope.

    \item \textbf{Andor at the 1.3-m McGraw-Hill Telescope (MDM):}\\
    The 1.3-m McGraw-Hill Telescope is located at the MDM Observatory, on the south-west ridge of Kitt Peak in Arizona. It was equipped with Andor Ikon DU937\_BV CCD camera, which was used in Frame Transfer mode and 4$\times$4 binning. We used a BG38 filter. One star was observed with this telescope.
 
    \item \textbf{SHOC at the SAAO 1.9-m Telescope and 1.0-m Telescope (SA19, SA10):} \\
    The telescopes are located at the Sutherland station of the South African Astronomical Observatory (SAAO), and are equipped with one of the Sutherland High Speed Optical Cameras (SHOC; \citealt{2013PASP..125..976C}). SHOC 1 and 2 are high-speed cameras operating in frame-transfer mode for visible wavelength range that have an electron-multiplying (EM) capability\footnote{The EM mode has not been used for observations presented in this work.}. The imaging area of the detectors is 1024$\times$1024 pixels, which corresponds to a FOV of 2.79$\times$2.79 arcmin for the 1.9-m telescope with the focal reducer, and 2.85$\times$2.85 arcmin for the 1.0-m telescope. A selection of amplifiers can be used, each resulting in a different gain setting, as well as binning and readout speed. The slowest readout speed was usually chosen, resulting in the lowest readout noise. Binning was determined by the observer to match the observing conditions and especially avoid undersampling of the Point Spread Function (PSF). Observations were done without a filter. Four stars were observed with these telescopes.  
\end{itemize}

The data were reduced using the following procedures. For data from DK we applied standard IRAF routines for all reduction steps. We extracted bad columns and hot pixels from the night's bias frames and flat fields, then cleaned the images for bad and hot pixels after the basic reduction steps (bias subtraction, dark and flat correction). As the last step, we checked for intensity gradients in the x and y directions (which sometimes occur in the presence of a bright Moon) and removed them, if necessary. The data from all the other instruments were reduced using standard \texttt{Astropy} \citep{2013AA...558A..33A,2018AJ....156..123A,2022ApJ...935..167A} \texttt{ccdproc} \citep{matt_craig_2017_1069648} routines consisting of bias subtraction, dark correction (only for observations with ProEM), flat-field and gain correction. Then, we performed aperture photometry using our own photometry pipeline with the use of adaptive circular apertures with sizes scaled to the seeing conditions for each frame \citep{2018MNRAS.479.2476S, 2021ApJ...918L...1S} with a scaling factor determined for each star and run. Comparison stars were chosen (wherever possible) such that they were brighter than the target and close to it, isolated and outside any faint nebulae, and when the target was the brightest in the field, an ``artificial'' comparison star comprising the summed flux from up to three available comparison stars was used. Because our target stars usually are much hotter than the available comparison stars, the differential light curves were corrected for differential color extinction by fitting a straight line to a Bouguer plot (differential magnitude vs. air mass). 
In the final step, we cleaned the light curves by removing outliers ($3.5\sigma$ clipping) and parts of data with bad quality (e.g., observations through thick clouds). We also inspected our differential magnitudes plotted against FWHM measurements to make sure that there is no correlation introduced by our photometry procedure. The constancy of the comparison stars was checked by examining differential light curves when more than one comparison star could be used. In case of fields with only a single comparison star, we looked up their {\it Gaia} $G_{\mathrm{BP}}-G_{\mathrm{RP}}$ colors, transformed these to $V-I_c$\footnote{\url{https://gea.esac.esa.int/archive/documentation/GDR3/Data_processing/chap_cu5pho/cu5pho_sec_photSystem/cu5pho_ssec_photRelations.html}}, and those to $B-V$ \citep{1993SAAOC..15....1C}. In that way, and with a rough correction for interstellar reddening, we inferred that none of the single comparison stars had $(B-V)_0<0.7$ and hence none of them lies in a $\kappa$-driven instability strip.

In this work, we present the results for a sample of 29 PG~1159 stars that are not surrounded by bright planetary nebulae. The list of targets, observing log and information on the scaling factor used in the photometry procedure is given in Table~\ref{tab:obs_log}. The light curves are presented in Figure~\ref{fig:survey1}.

\centerwidetable
\begin{deluxetable*}{lllclrclDc}
    \tablewidth{\textwidth} 
    \tabletypesize{\footnotesize}
\tablecaption{Log of photometric observations \label{tab:obs_log}}
\tablehead{\colhead{Name} & \colhead{Equip.}& \colhead{Observer} & \colhead{Date} & \colhead{Filter} & \colhead{$t_{\mathrm{exp}}$} & \colhead{Scale} & \colhead{$\Delta T$} & \multicolumn2c{$\Delta f$} & \colhead{Med. noise} \\
\colhead{} & \colhead{}& \colhead{} & \colhead{+UTC Start} & \colhead{} & \colhead{(s)} & \colhead{factor} & \colhead{} & \multicolumn2c{(d$^{-1}$)} & \colhead{(mmag)}
} 
\decimals
 \startdata
BMP 0739$-$1418 & DK & EP & 2014-12-26T05:48:30 & no filter & 30 & 1.5 & 2.39 hr & 10.06 & 0.31 \\ \hline
H1504+65 & GTC & SA & 2016-03-09T02:07:53 & Sloan r & 6 & 1.5 & 48 min & 30.00 & 0.42 \\ \hline
HS 0444+0453 & DK & EP & 2014-12-26T03:44:44 & no filter & 20 & 1.2 & 1.81 hr & 13.24 & 0.52 \\ \hline
HS 0704+6153 & GTC & SA & 2016-03-09T22:30:36 & Sloan r & 10 & 1.5 & 48 min & 30.34 & 0.53 \\ \hline
HS 1517+7403 & MD & GH & 2016-05-24T02:59:58 & BG40 & 10 & 0.9 & 1.51 hr & 15.88 & 0.61 \\ \hline
MCT 0130$-$1937 & SA19 & PS & 2014-12-05T19:15:59 & no filter & 10 & 0.9 & 2.30 hr & 10.45 & 0.83 \\ \hline
PG 1151$-$029 & INT & NH & 2016-03-29T21:44:06 & Harris V & 10 & 1.5 & 1.45 hr & 16.61 & 0.57 \\ \hline
PG 1520+525 & MD& GH & 2016-05-30T02:47:51 & BG40 & 15 & 1.2 & 1.15 hr & 20.87 & 0.66 \\ \hline
\multirow{2}{*}{PN A66 (Abell) 21} & DK & EP & 2015-02-10T02:01:46 & no filter & 40 & 1.2 & 1.16 hr & 20.71 & 2.02 \\ 
& GTC & SA & 2016-03-08T22:26:12 & Sloan r & 6 & 1.2 & 50 min & 29.04 & 0.44 \\ \hline
\multirow{2}{*}{PN A66 (Abell) 72} & \multirow{2}{*}{SA10} & \multirow{2}{*}{FW} & 2022-10-07T18:35:41 & no filter & 25 & 1.8 & 3.05 hr & 7.87 & 1.29 \\ 
 &  &  & 2022-10-08T18:01:35 & no filter & 30-35 & 1.5 & 2.04 hr & 11.76 & 1.58 \\ \hline
PN IsWe 1 & INT & MK+Students & 2016-10-19T04:11:53 & Harris V & 5 & 0.9 & 1.94 hr & 12.40 & 0.81 \\ \hline
\multirow{5}{*}{PN Jn 1} & INT & LSA, PSh & 2016-12-12T20:47:54 & Harris V & 10 & 1.2 & 3.01 hr & 7.98 & 0.81 \\ 
 & MD & JC & 2017-08-16T07:54:09 & BG40 & 10 & 1.2 & 3.75 hr & 6.40 & 0.30 \\ 
 & MD & JC & 2017-08-17T09:37:05 & BG40 & 10 & 1.2 & 1.93 hr & 12.41 & 0.35 \\ 
 & INT & DJ & 2017-08-28T01:45:39 & Harris V & 5 & 1.5 & 3.71 hr & 6.47 & 0.41 \\ 
 & INT & DJ & 2017-08-30T03:03:13 & Harris V & 10 & 1.2 & 2.64 hr & 9.09 & 0.45 \\ \hline
PN Lo (Longmore) 3 & DK & EP & 2015-02-10T00:41:13 & no filter & 40 & 1.2 & 1.16 hr & 20.65 & 2.40 \\ \hline
\multirow{2}{*}{RX J0122.9$-$7521} & \multirow{2}{*}{SA19} & \multirow{2}{*}{PS} & 2014-12-04T19:05:30 & no filter & 10 & 0.9 & 1.95 hr & 12.32 & 0.49 \\ 
 &  &  & 2014-12-09T18:32:43 & no filter & 10 & 0.9 & 2.42 hr & 9.92 & 0.46 \\ \hline
SDSS J000945.46+135814.4 & GTC & SA & 2017-12-06T22:18:56 & no filter & 10 & 1.2 & 58 min & 24.63 & 2.55 \\ \hline
SDSS J001651.42$-$011329.3 & SA19 & PS & 2017-12-06T22:18:55 & Sloan r & 20 & 0.9 & 1.71 hr & 14.04 & 1.58 \\ \hline
SDSS J055905.02+633448.4 & GTC & SA & 2017-09-15T04:09:52 & Sloan r & 20 & 1.2 & 59 min & 24.43 & 0.98 \\ \hline
SDSS J075540.94+400918.0 & GTC & SA & 2016-03-06T23:31:33 & Sloan r & 15 & 1.2 & 57 min & 25.21 & 0.59 \\ \hline
\multirow{3}{*}{SDSS J093546.53+110529.0} & DK & EP & 2015-01-02T06:26:02 & no filter & 30 & 0.9 & 1.55 hr & 15.45 & 1.37 \\ 
 & MD & JC & 2017-05-06T03:00:17 & BG40 & 30 & 1.2 & 2.92 hr & 8.23 & 1.85 \\ 
 & GTC & SA & 2018-08-14T14:45:58 & Sloan r & 20 & 1.2 & 57 min & 25.05 & 0.86 \\ \hline
SDSS J102327.41+535258.7 & INT & LPA & 2016-02-03T02:40:54 & Harris V & 20 & 1.2 & 2.52 hr & 9.54 & 1.58 \\ \hline
\multirow{6}{*}{SDSS J105300.24+174932.9} & MD & JC & 2017-05-02T03:34:16 & BG40 & 20 & 0.9 & 2.27 hr & 10.56 & 0.78 \\ \
 & MD & JC & 2017-05-07T03:08:52 & BG40 & 22 & 0.9 & 3.04 hr & 7.89 & 1.91 \\ 
 & GTC & SA & 2017-12-29T02:27:34 & Sloan r & 10 & 1.2 & 1.05 hr & 22.96 & 0.57 \\ 
 & MDM & KB & 2019-04-24T04:19:17 & BG38 & 30 & 1.5 & 1.99 hr & 12.09  & 0.79  \\ 
 & MDM & KB & 2019-04-25T03:36:34 & BG38 & 30 & 1.2 & 1.82 hr    & 13.20  & 1.55  \\ 
 & MDM & KB & 2019-04-26T02:54:19 & BG38 & 30 & 1.2 & 4.00 hr & 6.01 & 1.11 \\ \hline
\multirow{2}{*}{SDSS J121523.09+120300.8} & DK & EP & 2015-04-14T02:32:52 & no filter & 40 & 0.9 & 1.26 hr & 19.04 & 2.97 \\
& GTC & SA & 2018-01-17T06:01:53 & Sloan r & 20 & 1.2 & 1.13 hr & 21.17 & 0.95 \\ \hline
SDSS J123930.61+244321.7 & INT & PS, MT & 2016-03-11T00:46:00 & Harris V & 20 & 0.9 & 2.10 hr & 11.42 & 1.24 \\ \hline
\multirow{2}{*}{SDSS J134341.88+670154.5} & \multirow{2}{*}{MD} & \multirow{2}{*}{GH} & 2016-05-26T02:57:55 & BG40 & 20 & 1.5 & 59 min & 24.56 & 1.11 \\ 
 &  &  & 2016-05-29T02:47:32 & BG40 & 20 & 1.2 & 1.50 hr & 16.00 & 0.92 \\ \hline
SDSS J141556.26+061822.5 & MD & JC & 2017-05-05T05:14:23 & BG40 & 30 & 1.2 & 5.20 hr & 4.62 & 0.69 \\ \hline
SDSS J144734.12+572053.1 & MD & GH & 2016-05-28T02:46:07 & BG40 & 30 & 1.2 & 1.59 hr & 15.08 & 2.25 \\ \hline
SDSS J191845.01+624343.7 & MD & JC & 2017-05-08T07:08:02 & BG40 & 30 & 0.9 & 4.15 hr & 5.78 & 1.05 \\ \hline
Sh 2$-$68 & GTC & SA & 2016-04-23T04:45:44  & Sloan r & 10  & 1.5 & 58 min & 24.81 & 0.58 \\ \hline
Sh 2$-$78 & GTC & SA & 2016-04-24T04:27:31  & Sloan r & 15 & 1.2 & 1.24 hr & 19.30 & 0.75 
\enddata
\tablecomments{SA -- Support Astronomer; Students -- Rosa Clavero, Francisco Galindo, Bartosz Gauza; GTC -- GTC+OSIRIS; DK -- DK+DFOSC; MD -- McDonald 2.1-m+ProEM; MDM -- MDM 1.3-m+Andor; SA19 -- SAAO 1.9-m+SHOC; SA10 -- SAAO 1.0-m+SHOC; INT -- INT+WFC.\\ We refer to the central stars using the PN designations throughout the paper.}
\end{deluxetable*}


\section{Frequency analysis}
The light curves prepared in the previous step were the subject of frequency analysis. We used \texttt{Period04} \citep{2005CoAst.146...53L} to calculate Fourier amplitude spectra for each star and run separately, up to the corresponding Nyquist frequency. The Fourier amplitude spectra are shown in Figure~\ref{fig:survey1}. The frequency range for which our survey is sensitive to varies from star to star. The length of observations varied from slightly below an hour to a few hours, resulting in poor frequency resolution for the shortest ones (based on \citet{1978Ap&SS..56..285L} criterion of $\Delta{f} = 1.0/\Delta{T}$ for only the detection of modes\footnote{We note that for a correct determination of amplitudes and phases the criterion is  $\Delta{f} > 1.5/\Delta{T}$.}). For each Fourier amplitude spectrum, we calculated the median noise level, as well as our detection threshold (dashed line in Fig.~\ref{fig:survey1}), adopted as amplitude ratio of $S/N\geq{4}$ \citep{1993AA...271..482B}. Table~\ref{tab:obs_log} includes the length of observations, corresponding frequency resolution, and median noise level in the Fourier spectra for all observed targets. 

\begin{figure*}
	\includegraphics[width=\textwidth]{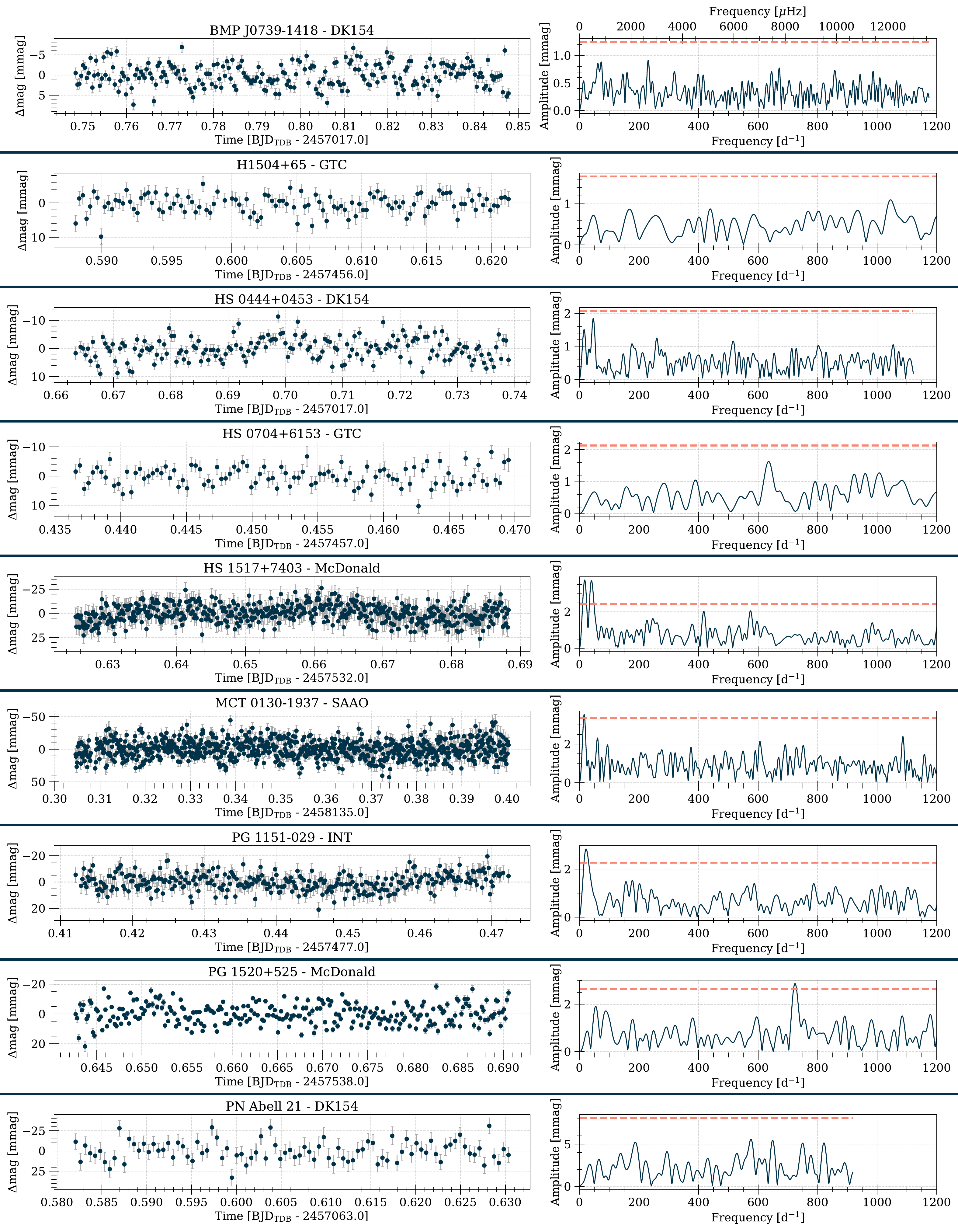}
    \caption{Light curves and their respective Fourier amplitude spectra of the survey targets. Plots for different stars are separated with horizontal lines. \textit{Light curves:} Note different scales. \textit{Fourier spectra:} they were calculated up to their respective Nyquist frequencies, but are plotted until 1200~d$^{-1}$. Dashed lines show the detection threshold of S/N$\geq{4}$. Note different scales.}
    \label{fig:survey1}
\end{figure*}

\setcounter{figure}{1}

\begin{figure*}
	\includegraphics[width=\textwidth]{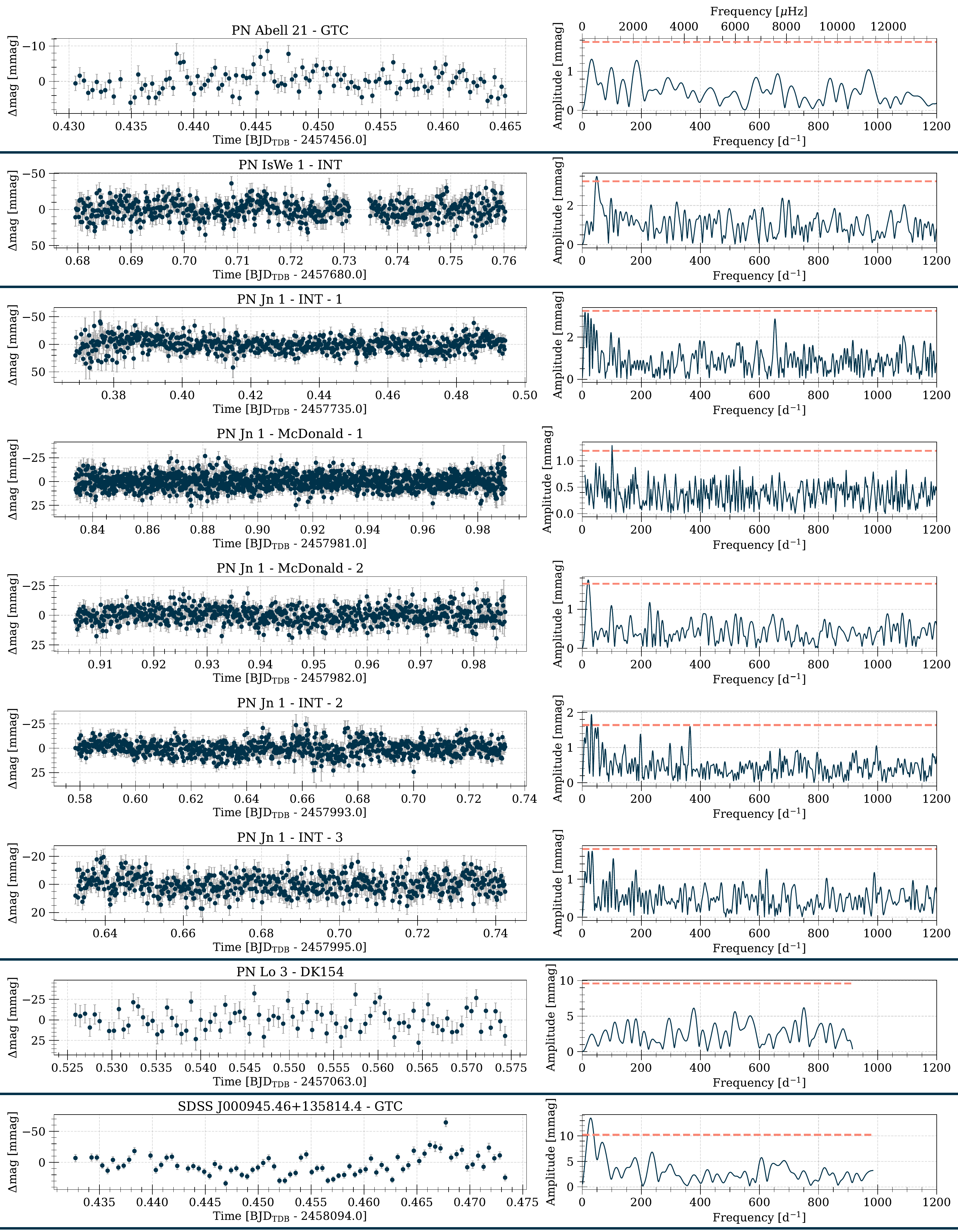}
    \caption{Continued.}
\end{figure*}

\setcounter{figure}{1}

\begin{figure*}
	\includegraphics[width=\textwidth]{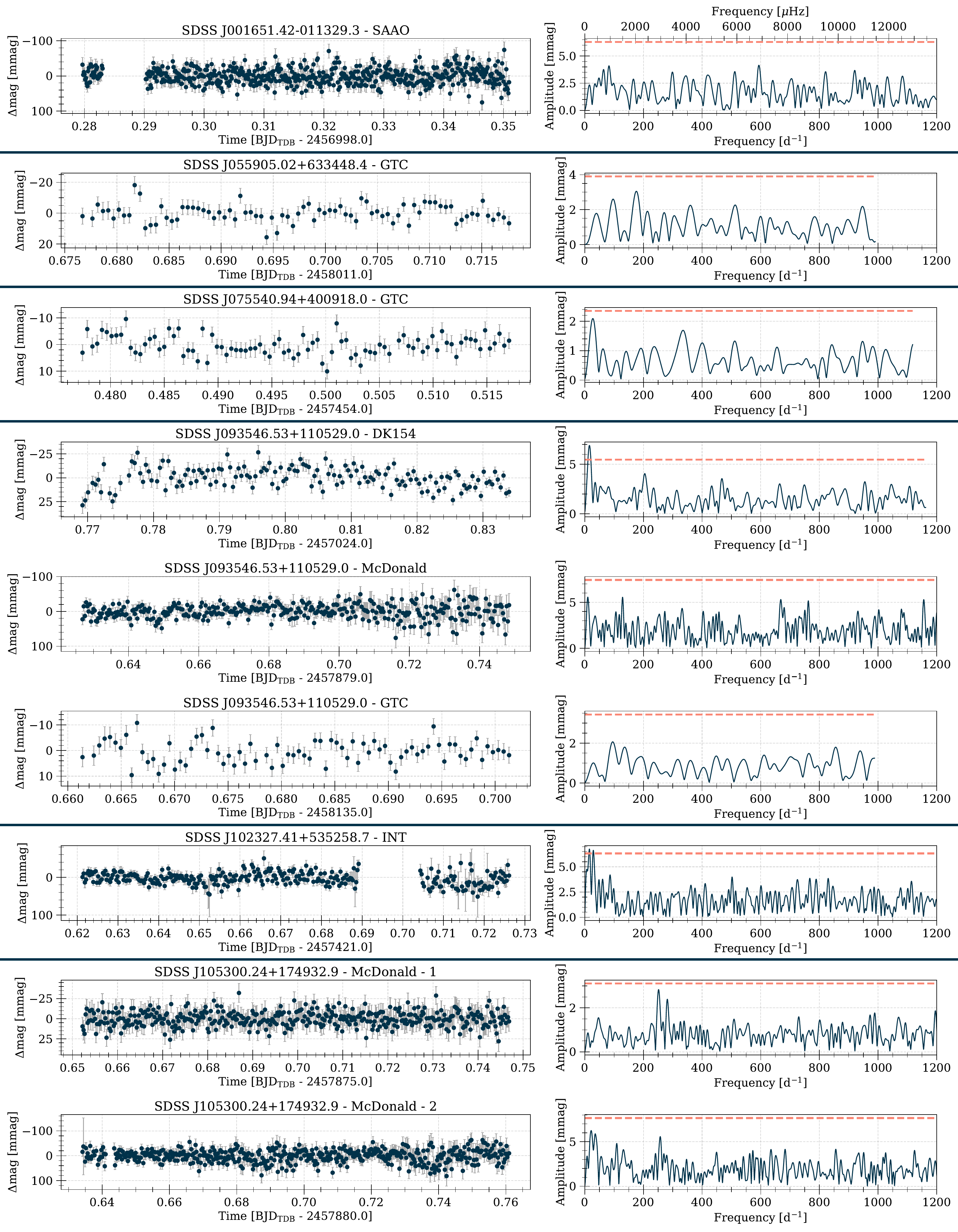}
    \caption{Continued.}
\end{figure*}

\setcounter{figure}{1}

\begin{figure*}
	\includegraphics[width=\textwidth]{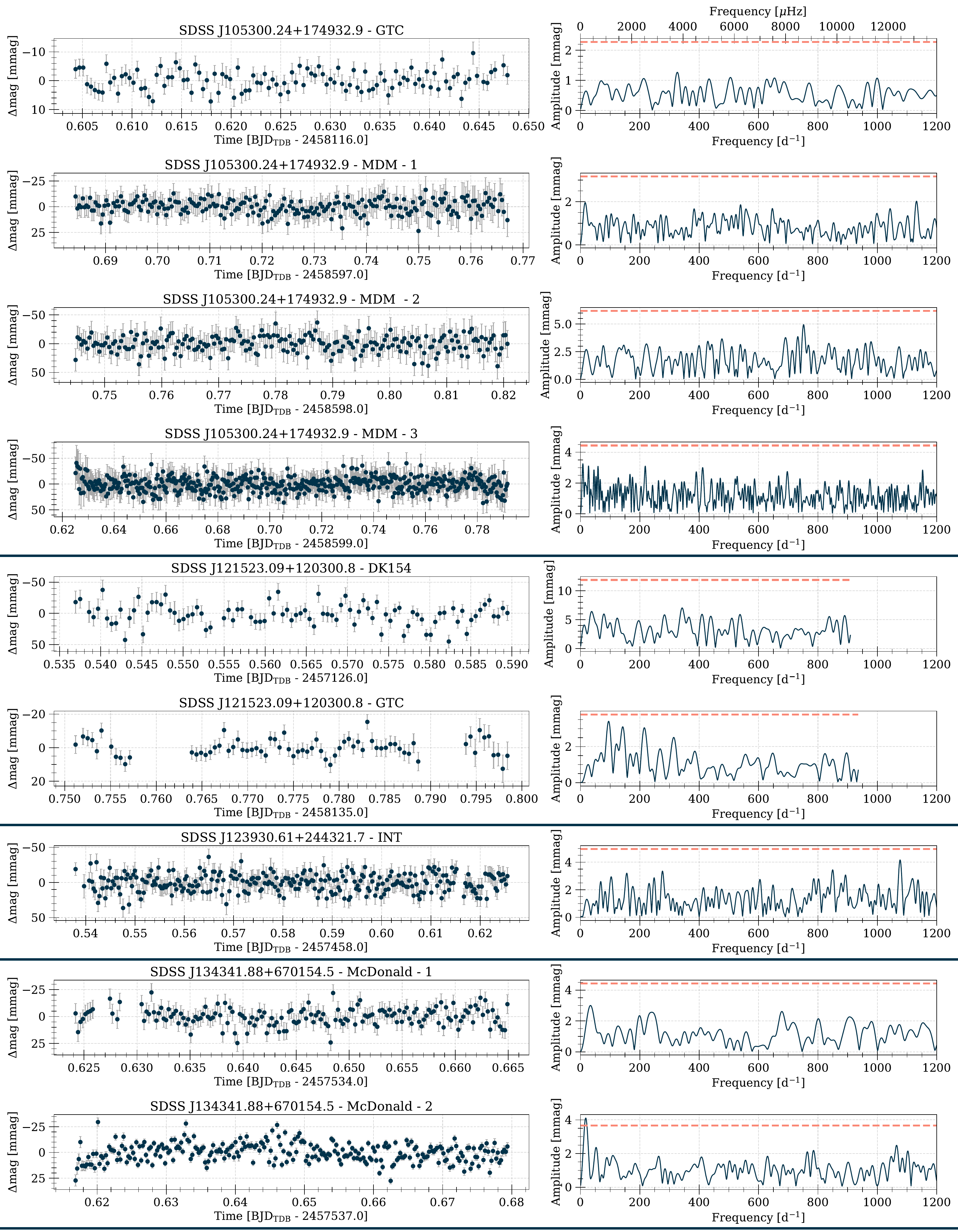}
    \caption{Continued.}
\end{figure*}

\setcounter{figure}{1}

\begin{figure*}
	\includegraphics[width=\textwidth]{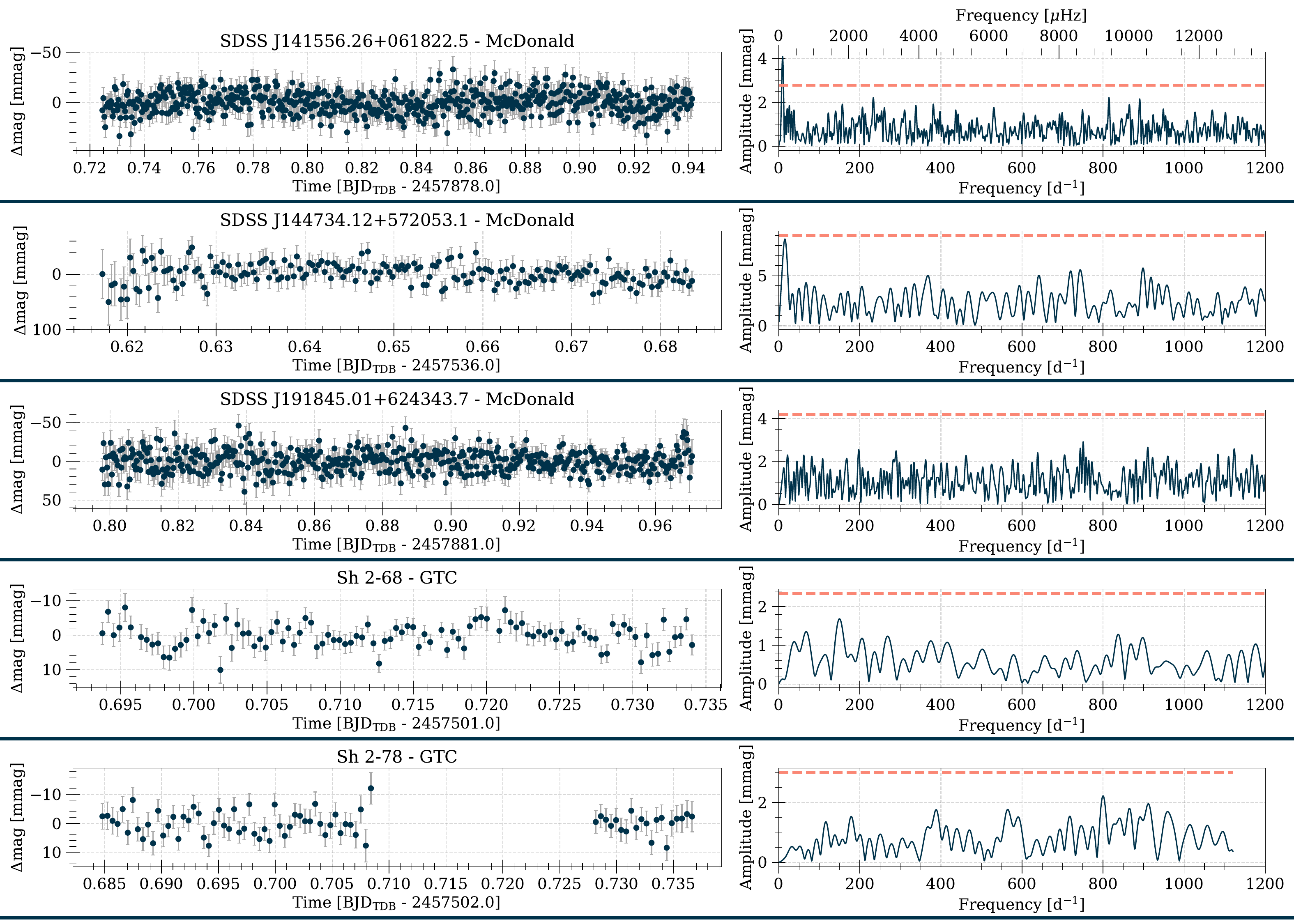}
    \caption{Continued.}
\end{figure*}


\section{Survey results}
The bottom panel of Figure~\ref{fig:hist} presents a histogram of the number of stars vs. the median noise level in the Fourier spectrum. In cases when the same star was observed multiple times, the lowest achieved level was taken. For the majority of 
observed stars, we reached a noise level of about 1 mmag or below. \citet{1987ApJ...323..271G} and \citet{1996AJ....111.2332C} reported their threshold for non-variable targets as the maximum amplitude in the Fourier spectra and reached values of $2.4-2.7$ and $2.4-5.3$ mmag, respectively. Inspection of Figure~\ref{fig:survey1} shows that our results are comparable to theirs, while our sample covered fainter stars ($15.4-18.6$ mag in $Gaia$, see the top panel of Fig.~\ref{fig:hist}). This allowed us to discover pulsations in the central star of planetary nebula Abell 72, and variability in RX J0122.9$-$7521. In addition to that, five objects from our sample could also be variable, but need follow-up observations for eventual confirmation. The majority of our sample did not show any variability consistent with GW Vir pulsations, and in those cases we put upper limits on non-variability. Each Fourier spectrum was also inspected for the presence of short-period $\epsilon$-driven modes \citep{2009ApJ...701.1008C}. No sign of such modes with periods shorter than about $200$~s (frequencies above 400~\cd) was found in any of the stars.


\section{Comments on selected stars}
Below, we comment on stars that showed peaks of interest in the Fourier spectrum. While most of our observations turned out to be non-detections, we have to mention one caveat. Pulsating PG 1159 stars are known for their variable pulsation spectra, even on a night-to-night basis. This is often caused by the interference between closely spaced modes, which occasionally becomes destructive and pushes the amplitudes of the modes below the detection threshold. Possible nonlinear mode coupling could have the same effect \citep[e.g.,][]{2011AA...528A...5V}. Non-detections because of those reasons could be avoided by observing the targets on multiple nights over the visibility period. While this was the case for eight targets, we were not able to acquire multiple runs for the remaining sample, and this has to be kept in mind regarding our non-detections.

\subsection{Pulsator -- PN Abell 72\,\footnote{We refer to the central stars using the PN designations throughout the paper.}}

\begin{figure*}
	\includegraphics[width=\textwidth]{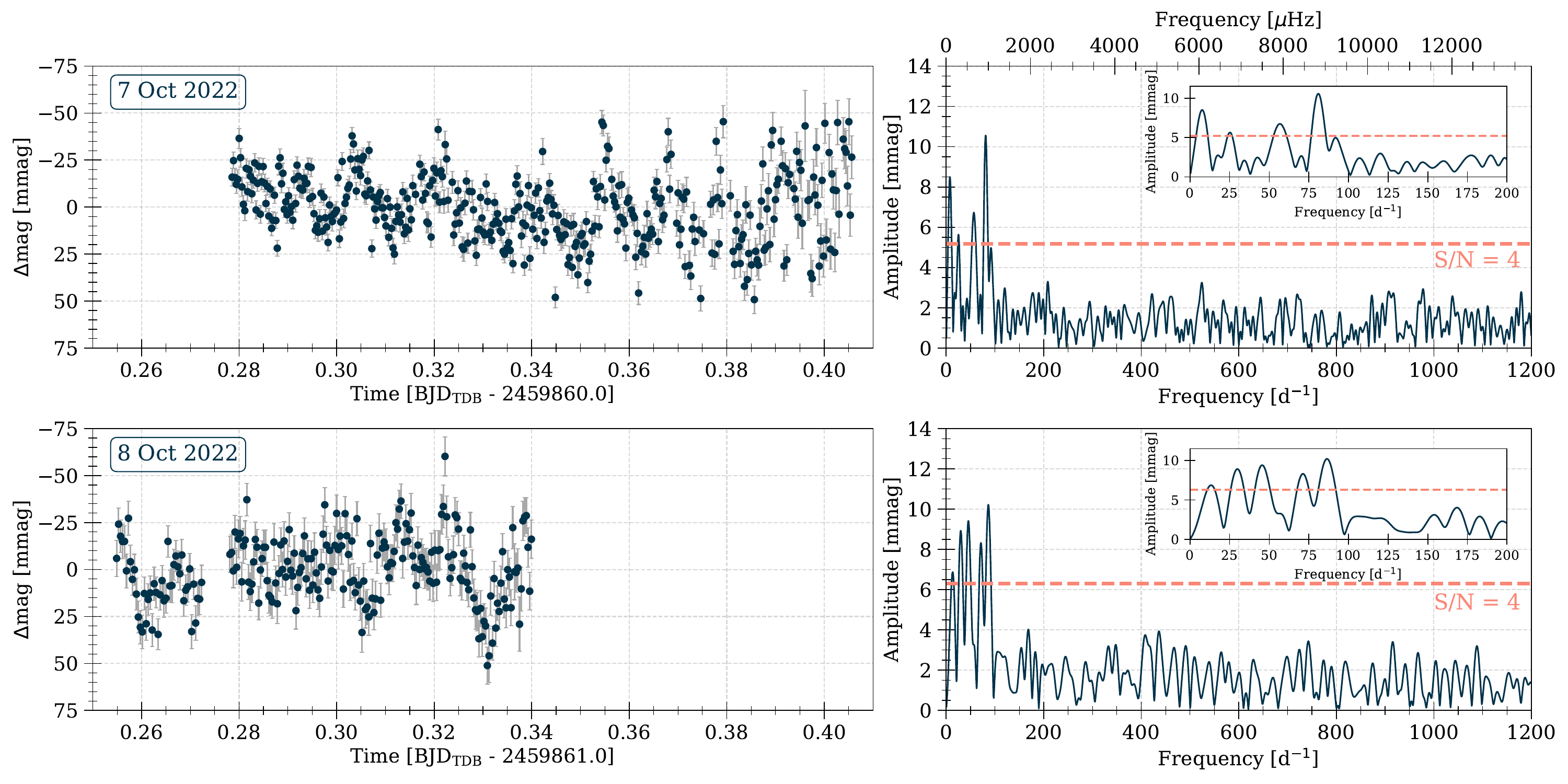}
    \caption{Light curves and Fourier amplitude spectra of two SAAO observing runs on the central star of planetary nebula Abell 72. The Fourier amplitude spectra were calculated up to the Nyquist frequency, but are shown up to 1200 d$^{-1}$. Insets show a zoomed in view into the frequency range of detected pulsations. Dashed lines show the detection threshold of S/N$\geq{4}$. Note the same scales for the light curves and Fourier spectra.}
    \label{fig:A72}
\end{figure*}

The central star of the planetary nebula Abell 72 was observed in October 2022 over two consecutive nights. The light curves and Fourier amplitude spectra are presented in Fig.~\ref{fig:A72}. We detected significant peaks reaching amplitudes on the order of 10 mmag in the nightly Fourier amplitude spectra, on both nights located in the same frequency range, consistent with g-mode pulsations seen in GW Vir stars. We classify Abell 72 as a multiperiodic pulsator and observations on a longer time base are needed to resolve its pulsation modes. 

\subsection{Candidates}\label{sec:cand}
\begin{itemize}

    \item \textbf{HS 0444+0453} \\
    There is an interesting, but statistically insignificant ($S/N=3.3$) peak around 45.7~d$^{-1}$ (period of about 1890~s). If confirmed, it fits within the observed period range of GW Vir pulsators.

    \item \textbf{HS 1517+7403} \\
    There are two statistically significant peaks: 17.5 and 40.5~\cd (S/N = 5.8 for both, periods of 4945 and 2133~s, respectively). Such long periods are usually found in GW Vir central stars of planetary nebulae, but no nebula around HS 1517+7403 has been reported. Given the short duration of the single run available and that only a single comparison star could be used, it is not clear whether these peaks are due to pulsations of the target.

    \item \textbf{PN IsWe 1} \\
    The highest, possibly unresolved peak at 49~\cd has S/N=4.4 and corresponds to a period of about 30~min (1764~s).
    While such pulsation periods are observed in GW Vir stars, observations on a longer time base are necessary for confirmation.

    \item \textbf{PN Jn 1} \\
    \citet{1996AJ....111.2332C} observed the star twice, obtaining peaks of maximum amplitude of 2.4 and 4.0 mmag in the Fourier amplitude spectra. They did not detect significant peaks (reaching 99\% confidence level) but two candidates: 540.5 and 538.5 $\mu{}$Hz (46.70 and 46.53\,\cd, respectively), and as a result did not claim the detection of pulsations in the central star of planetary nebula Jn 1. \citet{2006AA...454..527G} observed the star once and did not find the peaks tentatively detected by \citet{1996AJ....111.2332C}, instead they found a barely significant peak at 2200 $\mu{}$Hz (190.1\,\cd). Nevertheless, they claimed discovery of pulsations on that basis, but called for more observations to confirm their findings. We observed Jn 1 a total of five times in three different runs, and achieved very good median noise levels of 0.30--0.81 mmag. In none of the runs we saw signs of peaks previously reported or strong peaks occurring in more than one of our own runs (e.g., a peak at $~100$~\cd with $S/N=4.3$ only in McDonald--1 run). We thus conclude that there is no convincing evidence that Jn 1 pulsates, and that it requires observations of similar quality to our first McDonald run for eventual confirmation.

    \item \textbf{RX J0122.9$-$7521} \\
    RX J0122.9$-$7521 was observed twice in December 2014. The light curves and Fourier amplitude spectra are presented in Fig.~\ref{fig:RXJ01SAAO}. We detected a significant peak in the Fourier amplitude spectra of both nights, located at the same frequency of about 35~d$^{-1}$ and reaching an amplitude of $4-5$ mmag. RX J0122.9$-$7521 was also observed by \textit{TESS} in Sectors 1, 13, 27, and 28. The same frequency as in our ground-based data is present in the \textit{TESS} observations (34.78\,\cd). With $T_\mathrm{eff}=180 000$~K that would make RX J0122.9$-$7521 the hottest known variable/pulsating PG~1159 star. We further discuss this star in Sect.~\ref{rxj0122}.

\begin{figure*}
	\includegraphics[width=\textwidth]{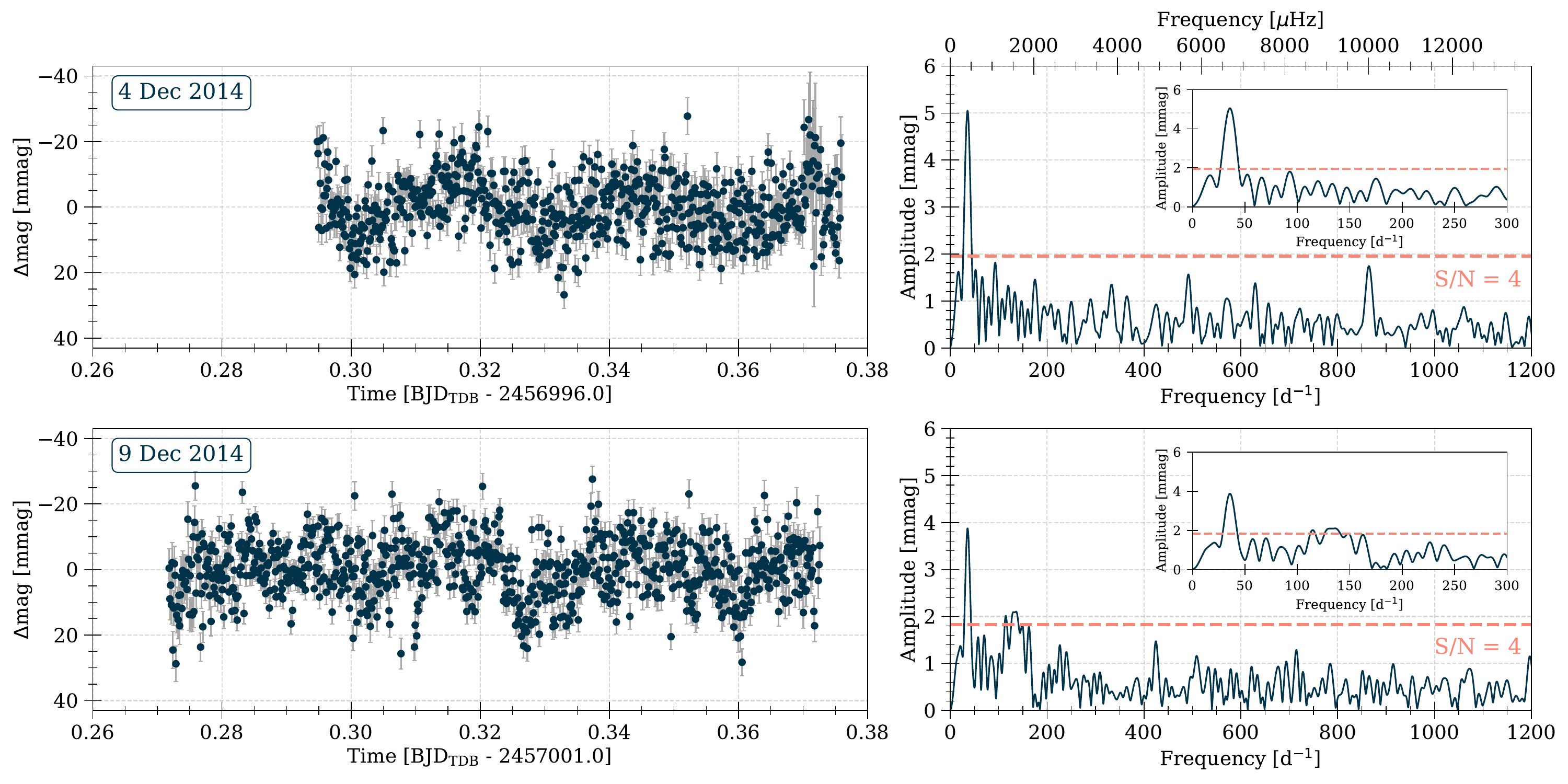}
    \caption{Light curves and Fourier amplitude spectra of two SAAO observing runs on RX J0122.9$-$7521. The Fourier amplitude spectra were calculated up to the Nyquist frequency, but are shown up to 1200 d$^{-1}$. Insets show a zoomed in view into the frequency range of the detected variations. Dashed lines show the detection threshold of S/N$\geq{4}$. Note the same scales for the light curves and Fourier spectra.}
    \label{fig:RXJ01SAAO}
\end{figure*}

    \item \textbf{SDSS J102327.41+535258.7} \\
    One suitable comparison star was used. There are two peaks: 15.7 and 30.4~\cd with $S/N$ of 4.3 and 4.2, which correspond to periods of 92 and 47~min (5507 and 2839~s), respectively.

    \item \textbf{SDSS J105300.24+174932.9} \\
    There are two interesting, but insignificant ($S/N=3.7$ and 3.1) peaks at 251.9 and 281.5~\cd, respectively, in the first McDonald run. In the second run, five days later, a peak in the same frequency region is present (at about 257~\cd), but due to higher noise the signal (if real) has only slightly higher amplitude than the highest noise peaks. In the remaining runs we were not able to reach a better noise level than in the first McDonald run, except the GTC run (that was too short).
\end{itemize}

\subsection{Nonpulsators}
\begin{itemize}
    \item \textbf{MCT 0130$-$1937} \\
    There is a significant low-frequency trend in the light curve (around 10~d$^{-1}$) that is likely not intrinsic to the star, particularly because there was only one comparison star available that was fainter than the target.

    \item \textbf{PG 1151$-$029} \\
    There is a significant low-frequency trend in the light curve, corresponding to a peak around 20~d$^{-1}$), that is likely caused by sky transparency variations that night.

    \item \textbf{PG 1520+525} \\
    There is a significant peak ($S/N=4.33$) at 725~d$^{-1}$. To assess whether this peak is real, we calculated differential light curves between the target and two different comparison stars, as well as between these comparison stars, and then computed the FTs of these light curves. The aforementioned peak showed up only in the difference of (target$-$comparison star 1), and not, as expected if the target was variable, also in the difference of (target$-$comparison star 2). We conclude that, even though formally significant in that differential light curve, this peak is not intrinsic to PG 1520+525.

    \item \textbf{SDSS J000945.46+135814.4} \\
    The observing run is too short (hence too low frequency resolution) to decide whether the signals around 50~d$^{-1}$ may be intrinsic to the star. 

    \item \textbf{SDSS J093546.53+110529.0} \\
    In the FT from DK there is a peak around 16~d$^{-1}$ that is due to a low-frequency trend in the light curve and not present in the runs from the other two instruments, one of which has a much lower noise level. We therefore conclude that this signal is not due to pulsations from the target.

    \item \textbf{SDSS J134341.88+670154.5} \\
    The second run shows a significant low-frequency trend around 20~d$^{-1}$ that is likely due to variable sky conditions and not intrinsic to the star. 

    \item \textbf{SDSS J141556.26+061822.5} \\ 
    There is a low-frequency peak at 9.46~\cd with $S/N=5.8$. This peak corresponds to a period of about 2.5~h (9000~s), which is too long for GW Vir pulsations.  
    This peak is also present in differential light curves between the target and either of the comparison stars. Therefore, the peak might be intrinsic to the target, but of different origin than GW Vir pulsations, e.g., rotation, binarity, or spots.
\end{itemize}


\pagebreak

\section{Impurity of GW Vir instability strip}

Previous observations showed that only about 50\% of stars within the GW Vir instability strip pulsate (see, e.g., Fig. 2 in \citealt{2022MNRAS.513.2285U}). A significant number of PG 1159 stars were discovered since then, and with our new results, we can re-determine the current occurrence rate for PG 1159 stars. In Table~\ref{tab:PG1159_physical} we listed the physical parameters of all known PG 1159 stars with updated information about their variability. For a total of 67 PG~1159 stars, 24 stars are confirmed as pulsating, which corresponds to 36\%. Still, the majority of PG~1159 stars within the instability strip are found to be non-pulsators. \citet{2021ApJ...918L...1S} recently showed that there was a clear separation between N-rich ($\approx 1\%$ N/He) pulsators and N-poor ($< 0.01\%$ N/He) non-pulsators. We therefore also listed the N abundance, where available, in Table~\ref{tab:PG1159_physical}. To date, only 26 PG~1159 stars have published N abundances.

\startlongtable
\begin{deluxetable*}{lrccllccl}
\tablecaption{Properties of PG 1159 stars \label{tab:PG1159_physical}}
\tabletypesize{\scriptsize}   
\tablehead{
 \colhead{Name} & \colhead{$T_{\mathrm{eff}}$}& \colhead{$\log{g}$} & \colhead{PN} & \colhead{Puls.} & \colhead{N} & \colhead{BC} & \colhead{$\log{L_{\star}}$} & \colhead{Ref.} \\ 
\colhead{} & \colhead{(K)}& \colhead{(cm s$^{-2}$)} & \colhead{} & \colhead{} & \colhead{} & \colhead{(mag)} & \colhead{L$_{\odot}$} & \colhead{}
} 
\startdata
    BMP J0739$-$1418 & 120000 & 6.0 & yes & \textbf{NOP} & N-poor & -7.269 & $3.48^{+0.11}_{-0.12}$ & W+2023 \\ \hline
    FEGU 248$-$5 & 160000 & 6.5 & yes & NVD & N-poor & -8.039 & $3.89^{+0.11}_{-0.10}$ & W+2023 \\ \hline    
    H1504+65         & 200000 & 8.0 & no & \textbf{NOP} & N-poor & -8.700 & $2.167^{+0.062}_{-0.064}$ & W+2004a, WD2005, WR2015, NW2004 \\ \hline
    HE 1429$-$1209     & 160000 & 6.0 & no & yes & no lit. data & -8.039 & $3.46^{+0.11}_{-0.12}$ & W+2004b \\ \hline
    HS 0444+0453     & 90000 & 7.0 & no & \textbf{NOP} & no lit. data & -6.429 & $1.28^{+0.14}_{-0.16}$ & D1999 \\ \hline
    HS 0704+6153     & 75000 & 7.0 & no & \textbf{NOP} & N-poor & -5.832 & $1.03^{+0.17}_{-0.20}$ & DH1998 \\ \hline
    HS 1517+7403     & 110000 & 7.0 & no & \textbf{NOP} & N-poor & -7.045 & $1.84^{+0.10}_{-0.12}$ & DH1998 \\ \hline
     HS 2324+3944     & 130000 & 6.2 & no & yes & no lit. data & -7.460 & $3.390^{+0.092}_{-0.084}$ & F+2010, S+1999, C+2021 \\ \hline
    MCT 0130$-$1937    & 90000 & 7.5 & no & \textbf{NOP} & N-poor & -6.429 & $1.42^{+0.14}_{-0.16}$ & W+2004c, WR2014 \\ \hline
    NGC 246          & 150000 & 5.7 & yes & yes & N-poor & -7.855 & $3.786^{+0.081}_{-0.084}$ & W+2005, CB1996 \\ \hline
    NGC 650          & 140000 & 7.0 & yes & NOP & no lit. data & -7.658 & $3.27^{+0.67}_{-0.36}$ & NS1995 \\ \hline
    NGC 6852         & 150000 & 6.0 & yes & yes & no lit. data & -7.855 & $2.93^{+0.34}_{-0.26}$ & K. Werner, GP+2006 \\ \hline
    NGC 7094         & 110000 & 5.7 & yes & yes & N-poor & -7.045 & $3.83^{+0.10}_{-0.12}$ & F+2010, S+2007 \\ \hline
    PG 0122+200      & 80000 & 7.5 & no & yes & N-rich & -6.043 & $1.20^{+0.17}_{-0.19}$ & WR2014, F+2007 \\ \hline
    PG 1144+005      & 150000 & 6.5 & no & yes & N-rich & -7.855 & $3.13^{+0.10}_{-0.10}$ & W+2005, W+2016, S+2021 \\ \hline
    PG 1151$-$029      & 140000 & 6.0 & no & \textbf{NOP} & no lit. data & -7.658 & $2.471^{+0.097}_{-0.091}$ & W+2004c \\ \hline
    PG 1159$-$035      & 140000 & 7.0 & no & yes & N-rich & -7.658 & $2.596^{+0.085}_{-0.086}$ & W+2005, W+2016, C+2008, O+2022 \\ \hline
    PG 1424+535      & 110000 & 7.0 & no & NOP & N-poor & -7.045 & $1.838^{+0.092}_{-0.115}$ & W+2005, W+2015 \\ \hline
    PG 1520+525      & 150000 & 7.5 & yes & \textbf{NOP} & N-poor & -7.866 & $2.591^{+0.081}_{-0.087}$ & W+2005, W+2016 \\ \hline
    PG 1707+427      & 85000 & 7.5 & no & yes & N-rich & -6.243 & $1.47^{+0.15}_{-0.17}$ & W+2005, W+2015, H+2018, K+2004 \\ \hline
    PN A66 21        & 140000 & 6.5 & yes & \textbf{NOP} & no lit. data & -7.658 & $2.118^{+0.088}_{-0.086}$ & W+2004c \\ \hline
    PN A66 43        & 110000 & 5.7 & yes & yes & N-rich & -7.045 & $3.69^{+0.10}_{-0.12}$ & F+2010, V+2005 \\ \hline
    PN A66 72        & 170000 & 6.5 & yes & \textbf{yes} & N-rich & -8.212 & $3.35^{+0.11}_{-0.12}$ & B+2023 \\ \hline
    PN IsWe 1        & 90000 & 7.0 & yes & \textbf{NOP} & no lit. data & -6.429 & $1.34^{+0.14}_{-0.16}$ & D1999 \\ \hline
    PN Jn 1          & 150000 & 6.5 & yes & \textbf{NOP} & no lit. data & -7.855 & $2.687^{+0.097}_{-0.095}$ & RW1995 \\ \hline
    PN K 1$-$16        & 160000 & 5.8 & yes & yes & no lit. data & -8.039 & $3.601^{+0.083}_{-0.088}$ & W+2010, G+1987, C+2021 \\ \hline
    PN Kn 12         & 170000 & 6.5 & yes & NVD & no lit. data & -8.212 & $3.20^{+0.36}_{-0.28}$ & B+2023\\ \hline
    PN Kn 61    & 170000 & 6.5 & yes & yes & N-rich & -8.212 & $3.54^{+0.37}_{-0.27}$ & DM+2015, B+2023, S+2023\\ \hline
    PN Kn 130        & 170000 & 6.5 & yes & NVD & N-poor & -8.212 & $3.40^{+0.13}_{-0.13}$ & B+2023\\ \hline
    PN Lo 3          & 140000 & 6.3 & yes & \textbf{NOP} & no lit. data & -7.658 & $3.08^{+0.15}_{-0.14}$ & W+2004c \\ \hline
    PN Lo 4          & 170000 & 6.0 & yes & yes & N-poor & -8.212 & $3.65^{+0.18}_{-0.13}$ & W+2010, BM1990 \\ \hline
    PN Ou 2	         & 170000 & 6.5 & yes & NVD & no lit. data & -8.212 & $2.28^{+0.46}_{-0.22}$ & B+2023 \\ \hline
    PN VV 47         & 130000 & 7.0 & yes & NOP & no lit. data & -7.460 & $2.04^{+0.11}_{-0.10}$ & RW1995 \\ \hline
    RL 104           & 80000 & 6.0 & no & NVD & N-rich & -6.046 & $3.17^{+0.15}_{-0.18}$ & W+2022 \\ \hline
    RX J0122.9$-$7521  & 180000 & 7.5 & no & \textbf{NOP} & no lit. data & -8.389 & $2.958^{+0.067}_{-0.071}$ & W+2004c \\ \hline
    RX J2117.1+3412  & 170000 & 6.0 & yes & yes & no lit. data & -8.212 & $3.394^{+0.067}_{-0.071}$ & W+2005, V+2002, C+2021 \\ \hline
    SALT J172411.7$-$632147 & 160000 & 6.5 & no & yes & N-poor & -8.039 & $3.01^{+0.11}_{-0.12}$ & J+2023 \\ \hline
    SALT J213742.6$-$382901 & 180000 & 7.0 & no & yes & N-rich & -8.376 & $3.04^{+0.17}_{-0.13}$ & J+2023 \\ \hline
    SDSS	J000945.46+135814.4 & 120000 & 7.5 & no & \textbf{NOP} & no lit. data & -7.279 & $2.49^{+0.44}_{-0.25}$ & K+2016 \\ \hline
    SDSS	J001651.42$-$011329.3 & 120000 & 5.5 & no & \textbf{NOP} & no lit. data & -7.269 & $3.19^{+0.22}_{-0.18}$ & H+2006 \\ \hline
    SDSS	J034917.41$-$005919.3 & 90000 & 7.5 & no & yes & no lit. data & -6.429 & $1.33^{+0.19}_{-0.17}$ & H+2006, W+2012 \\ \hline
    SDSS	J055905.02+633448.4 & 110000 & 7.5 & no & \textbf{NOP} & no lit. data & -7.050 & $1.49^{+0.17}_{-0.17}$ & W+2014 \\ \hline
    SDSS	J075415.11+085232.1 & 120000 & 7.0 & no & yes & no lit. data & -7.269 & $1.70^{+0.58}_{-0.29}$ & W+2014, K+2014 \\ \hline
    SDSS	J075540.94+400918.0 & 100000 & 7.6 & no & \textbf{NOP} & no lit. data & -6.764 & $1.62^{+0.19}_{-0.18}$ & H+2006 \\ \hline
    SDSS	J093546.53+110529.0 & 100000 & 7.6 & no & \textbf{NOP} & no lit. data & -6.764 & $1.47^{+0.17}_{-0.17}$ & H+2006 \\ \hline
    SDSS	J102327.41+535258.7 & 110000 & 7.6 & no & \textbf{NOP} & no lit. data & -7.050 & $2.15^{+0.23}_{-0.21}$ & H+2006 \\ \hline
    SDSS	J105300.24+174932.9 & 100000 & 7.0 & no & \textbf{NOP} & no lit. data & -6.762 & $1.59^{+0.13}_{-0.14}$ & W+2014 \\ \hline
    SDSS	J121523.09+120300.8 & 100000 & 7.6 & no & \textbf{NOP} & no lit. data & -6.764 & $1.65^{+0.25}_{-0.21}$ & H+2006 \\ \hline
    SDSS	J123930.61+244321.7 & 100000 & 7.5 & no & \textbf{NOP} & no lit. data & -6.764 & $1.64^{+0.25}_{-0.21}$ & W+2014 \\ \hline
    SDSS	J134341.88+670154.5 & 100000 & 7.6 & no & \textbf{NOP} & no lit. data & -6.764 & $1.44^{+0.12}_{-0.14}$ & H+2006 \\ \hline
    SDSS	J141556.26+061822.5 & 120000 & 7.5 & no & \textbf{NOP} & no lit. data & -7.279 & $1.83^{+0.15}_{-0.13}$ & W+2014 \\ \hline
    SDSS	J144734.12+572053.1 & 100000 & 7.6 & no & \textbf{NOP} & no lit. data & -6.764 & $1.58^{+0.19}_{-0.17}$ & H+2006 \\ \hline
    SDSS	J152116.00+251437.5 & 140000 & 6.0 & no & NOP & no lit. data & -7.658 & $3.11^{+0.42}_{-0.29}$ & W+2014 \\ \hline
    SDSS	J155610.40+254640.3 & 100000 & 5.3 & no & NVD & no lit. data & -6.762 & $3.23^{+0.40}_{-0.37}$ & R+2016 \\ \hline
    SDSS	J163727.03+485355.2 & 100000 & 7.5 & no & NVD & no lit. data & -6.764 & $1.86^{+0.39}_{-0.22}$ & K+2016 \\ \hline
    SDSS	J191845.01+624343.7 & 100000 & 7.2 & no & \textbf{NOP} & no lit. data & -6.762 & $1.65^{+0.14}_{-0.15}$ & W+2014 \\ \hline
    SDSS	J212531.92$-$010745.8 & 100000 & 7.5 & no & NOP & no lit. data & -6.764 & $2.54^{+0.41}_{-0.25}$ & K. Werner \\ \hline
    Sh	2$-$68 & 84000 & 7.2 & no & \textbf{NOP} & no lit. data & -6.205 & $1.70^{+0.17}_{-0.16}$ & G+2010 \\ \hline
    Sh	2$-$78 & 120000 & 7.5 & yes & \textbf{NOP} & no lit. data & -7.279 & $1.79^{+0.11}_{-0.11}$ & D1999 \\ \hline
    TIC	95332541 & 100000 & 7.5 & no & yes & N-poor & -6.764 & $2.14^{+0.12}_{-0.13}$ & U+2021, R+2023 \\ \hline
    TIC	333432673 & 120000 & 7.5 & no & yes & no lit. data & -7.279 & $1.924^{+0.082}_{-0.095}$ & U+2021 \\ \hline 
    TIC	403800675 & 110000 & 7.5 & no & yes & no lit. data & -7.050 & $1.73^{+0.10}_{-0.12}$ & U+2022 \\ \hline
    TIC	1989122424 & 110000 & 7.5 & no & yes & no lit. data & -7.050 & $1.29^{+0.11}_{-0.13}$ & U+2022 \\ \hline
    WD J070204.29+051420.56 & 100000 & 7.5 & no & NVD & N-poor & -6.764 & $1.63^{+0.12}_{-0.14 }$ & R+2023 \\ \hline \hline
    NGC	6765 & - & - & yes & NVD & no lit. data & - & - & NS1995 \\ \hline
    PG	2131+066 & 95000 & 7.5 & no & yes & N-rich & -  & - & WR2014, K+1995 \\ \hline
    RX	J0439.8$-$6809 & 250000 & 8.0 & no & NOP & N-poor & -  & - & WR2015
\enddata
\tablecomments{Properties of PG 1159 stars. \textbf{Bold}--this work. NOP -- Not Observed to Pulsate, NVD -- No Variability Data available. The last three stars either lack \textit{Gaia} measurements or T$_{\mathrm{eff}}$ and $\log{g}$ determinations and were excluded from the analysis. References: BM1990 -- \citet{1990AJ....100..788B},
B+2023 -- \citet{2023MNRAS.521..668B},
CB1996 -- \citet{1996AJ....111.2332C},
C+2008 -- \citet{2008AA...477..627C}, 
C+2021 -- \citet{2021AA...645A.117C}, 
DH1998 -- \citet{1998AA...334..618D}, 
D1999 -- \citet{1999RvMA...12..255D},
DM+2015 -- \citet{2015MNRAS.448.3587D},
F+2007 -- \citet{2007AA...467..237F}, 
F+2010 -- \citet{2010AIPC.1273..231F}, 
G+1987 -- \citet{1987fbs..conf..231G}, 
GP+2006 -- \citet{2006AA...454..527G}, 
G+2010 -- \citet{2010ApJ...720..581G}, 
H+2006 -- \citet{2006AA...454..617H}, 
H+2018 -- \citet{2018AA...612A..62H},
J+2023 -- \citet{2023MNRAS.519.2321J},
K+1995 -- \citet{1995ApJ...450..350K},
K+2004 -- \citet{2004AA...428..969K},
K+2014 -- \citet{2014MNRAS.442.2278K}, 
K+2016 -- \citet{2016MNRAS.455.3413K}, 
NS1995 -- \citet{1995AA...301..545N}, 
NW2004 -- \citet{2004AA...426L..45N}, 
O+2022 -- \citet{2022ApJ...936..187O}, 
RW1995 -- \citet{1995LNP...443..186R},
R+2016 -- \citet{2016AA...587A.101R},
R+2023 -- \citet{2023arXiv230703721R}, 
S+1999 -- \citet{1999AA...342..745S}, 
S+2007 -- \citet{2007AA...468.1057S}, 
S+2021 -- \citet{2021ApJ...918L...1S},
S+2023 -- Sowicka et al. in prep. (2023),
U+2021 -- \citet{2021AA...655A..27U}, 
U+2022 -- \citet{2022MNRAS.513.2285U},
V+2002 -- \citet{2002AA...381..122V},
V+2005 -- \citet{2005AA...433.1097V}, 
W+2004a -- \citet{2004AA...421.1169W},
W+2004b -- \citet{2004AA...424..657W}, 
W+2004c -- \citet{2004AA...427..685W},
W+2005 -- \citet{2005AA...433..641W}, 
WD2005 -- \citet{2005AA...434..707W}, 
W+2010 -- \citet{2010ApJ...719L..32W}, 
W+2012 -- \citet{2012MNRAS.426.2137W}, 
W+2014 -- \citet{2014AA...564A..53W}, 
WR2014 -- \citet{2014AA...569A..99W},  
W+2015 -- \citet{2015AA...582A..94W},
WR2015 -- \citet{2015AA...584A..19W}, 
W+2016 -- \citet{2016AA...593A.104W}, 
W+2022 -- \citet{2022AA...658A..66W},
W+2023 -- \citet{Weidmann2023}}
\end{deluxetable*}



\section{Properties of all known PG 1159 stars}
\label{s:PG1159properties}

Thanks to the \textit{Gaia} mission \citep{2016AA...595A...1G, 2023AA...674A...1G} the community received precise measurements of positions and distances of more than 1 billion stars. For the first time, consistent distance measurements became available for almost the entire sample of PG 1159 stars\footnote{With the exception of two stars without sufficient \textit{Gaia} data, which are listed at the bottom of Table~\ref{tab:astrometry}}. In Table~\ref{tab:astrometry} we compiled available \textit{Gaia} DR3 information for PG 1159 stars: identifiers, positions, \textit{Gaia} $G$ magnitudes and parallaxes with geometric distances determined by \citet{2021AJ....161..147B}. We also list the corresponding reddening E(B-V) at these distances, determined from the 3D reddening map of \citet{2018MNRAS.478..651G} (\texttt{Bayestar17}) using the Python package \texttt{dustmaps}. Even though a newer version of \texttt{Bayestar} is available (\texttt{Bayestar19}, \citealt{2019ApJ...887...93G}), it did not cover the distances of all the stars in our sample, hence we used the \texttt{Bayestar17} reddening map for all but six stars. Those six stars were not covered because of declination south of $-30^{\circ}$. For these cases, we used the 2D dust maps of \citet{2011ApJ...737..103S} and \citet{1998ApJ...500..525S} (SFD), which are equivalent to \texttt{Bayestar} in terms of units. We did not take into account the reddening by the surrounding planetary nebulae in the case of PG 1159 stars being the central stars of planetary nebulae. We also listed the \texttt{RUWE} (Renormalized Unit Weight Error) coefficient for each star, and marked in bold values higher than the canonical 1.4, which might either suggest an unreliable astrometric solution (in a few cases that corresponds with a large parallax error) or be a hint towards binarity. In the final column of Table~\ref{tab:astrometry} we put a remark for non-single stars  (e.g., known or suspected binaries/triples) and a subclass of so-called ``hybrid''-PG 1159 stars (exhibiting traces of hydrogen in the atmosphere).

\begin{longrotatetable}
\begin{deluxetable*}{lllDclcllll}
    \tabletypesize{\footnotesize}
    \tablecaption{Astrometric properties of PG 1159 stars\label{tab:astrometry}}
    \tablehead{\colhead{Name} & \colhead{\textit{Gaia} ID} & \colhead{RA} & \multicolumn2c{Dec.} & \colhead{\textit{Gaia} G} & \colhead{$\varpi{}_{\mathrm{Gaia}}$} & \colhead{$\sigma_{\varpi_{\mathrm{Gaia}}}$/$\varpi{}$} & \colhead{rgeo} & \colhead{E(B-V)} & \colhead{RUWE} & \colhead{Remarks} \\
\colhead{} & \colhead{} & \colhead{(deg)} & \multicolumn2c{(deg)} & \colhead{(mag)} & \colhead{(mas)} & \colhead{($\%$)} & \colhead{(kpc)} & \colhead{(mag)} & \colhead{} & \colhead{}
} 
\decimals
\startdata
        BMP J0739$-$1418 & 3030005560828868096 & 114.96064 & $-14.30718$ & 15.61 & $0.458 \pm 0.042$ & 9 & $2.10_{-0.19}^{+0.18}$ & $0.258 \pm 0.021$ & 1.041 &  \\ \hline	
        FEGU 248$-$5 & 5594969135329315072 & 115.59902 & $-32.79746$ & 17.00 & $0.528 \pm 0.052$ & 10 & $1.90_{-0.16}^{+0.19}$ & 0.944\tablenotemark{a} & 0.997 &  \\ \hline     
        H1504+65 & 1645296216119116928 & 225.54006 & +66.20535 & 16.29 & $2.156 \pm 0.051$ & 2 & $0.47_{-0.11}^{+0.12}$ & $0.0144 \pm 0.0028$ & 1.050 &  \\ \hline
        HE 1429$-$1209 & 6324298665725984512 & 218.08641 & $-12.38006$ & 16.01 & $0.441 \pm 0.054$ & 12 & $2.16_{-0.22}^{+0.23}$ & $0.101575 \pm 0.000099$ & 1.012 &  \\ \hline
        HS 0444+0453 & 3281864642080410112 & 071.76880 & +04.97804 & 16.23 & $2.271 \pm 0.062$ & 3 & $0.441_{-0.014}^{+0.011}$ & $0.0473 \pm 0.0033$ & 0.985 &  \\ \hline
        HS 0704+6153 & 1099093607199220096 & 107.38536 & +61.80533 & 16.98 & $1.643 \pm 0.074$ & 5 & $0.615_{-0.030}^{+0.033}$ & $0.0413 \pm 0.0050$ & 0.974 &  \\ \hline
        HS 1517+7403 & 1697669356564165632 & 229.19327 & +73.86865 & 16.63 & $1.319 \pm 0.061$ & 5 & $0.781_{-0.037}^{+0.031}$ & $0.0278 \pm 0.0024$ & 0.963 &  \\ \hline
        HS 2324+3944 & 1923253820774422272 & 351.81644 & +40.02323 & 14.77 & $0.702 \pm 0.034$ & 5 & $1.400_{-0.054}^{+0.074}$ & $0.1343 \pm 0.0013$ & 1.112 & hybrid \\ \hline
        MCT 0130$-$1937 & 5140121722033618560 & 023.16399 & $-19.36138$ & 15.76 & $2.395 \pm 0.066 $ & 3 & $0.414_{-0.011}^{+0.010}$ & $0.0297 \pm 0.0021$ & 1.283 &  \\ \hline
        NGC 246 & 2376592910265354368 & 011.76385 & $-11.87198$ & 11.80 & $1.799 \pm 0.079 $ & 4 & $0.538_{-0.017}^{+0.020}$ & $0.04481 \pm 0.00092$ & \textbf{1.530} & triple \\ \hline
        NGC 650 & 406328443354164480 & 025.58192 & +51.57541 & 17.42 & $0.294 \pm 0.203 $ & \textbf{69} & $3.7_{-1.5}^{+2.8}$ & $0.1431 \pm 0.0073$ & \textbf{1.727} &  \\ \hline
        NGC 6852 & 4237745794618477440 & 300.16337 & +01.72801 & 17.91 & $0.39 \pm 0.12$ & \textbf{30} & $3.0_{-0.9}^{+1.1}$ & $0.1083 \pm 0.0042$ & 1.017 &  \\ \hline
        NGC 7094 & 1770058865674512896 & 324.22072 & +12.78859 & 13.52 & $0.604 \pm 0.034$ & 6 & $1.607_{-0.076}^{+0.092}$ & $0.12600 \pm 0.00046$ & 0.970 & hybrid \\ \hline
        PG 0122+200 & 2786529465445503488 & 021.34385 & +20.29910 & 16.75 & $1.641 \pm 0.080$ & 5 & $0.618_{-0.032}^{+0.042}$ & $0.0396 \pm 0.0018$ & 0.982 &  \\ \hline
        PG 1144+005 & 3795664157996369024 & 176.64674 & +00.20928 & 15.16 & $0.802 \pm 0.058$ & 7 & $1.220_{-0.076}^{+0.085}$ & $0.02041 \pm 0.00080$ & 1.088 &  \\ \hline
        PG 1151$-$029 & 3601781534594624000 & 178.56280 & $-03.20143$ & 16.07 & $1.060 \pm 0.063$ & 6 & $0.938_{-0.046}^{+0.060}$ & $0.0382 \pm 0.0040$ & 1.044 &  \\ \hline
        PG 1159$-$035 & 3600841623951744640 & 180.44149 & $-03.76130$ & 14.69 & $1.691 \pm 0.064$ & 4 & $0.585_{-0.021}^{+0.020}$ & $0.0241 \pm 0.0031$ & 1.129 &  \\ \hline
        PG 1424+535 & 1605381435770077312 & 216.48109 & +53.25704 & 15.88 & $1.771 \pm 0.041$ & 2 & $0.566_{-0.011}^{+0.012}$ & $0.0126 \pm 0.0016$ & 1.033 &  \\ \hline
        PG 1520+525 & 1595941441250636672 & 230.44399 & +52.36779 & 15.55 & $1.295 \pm 0.041$ & 3 & $0.783_{-0.030}^{+0.027}$ & $0.0256 \pm 0.0029$ & 1.045 &  \\ \hline
        PG 1707+427 & 1355161726346266112 & 257.19864 & +42.68358 & 16.65 & $1.402 \pm 0.052$ & 4 & $0.733_{-0.026}^{+0.032}$ & $0.0477 \pm 0.0012$ & 1.002 &  \\ \hline
        PN A66 21 & 3163546505053645056 & 112.26128 & +13.24679 & 15.93 & $1.689 \pm 0.069$ & 4 & $0.584_{-0.021}^{+0.024}$ & $0.0318 \pm 0.0013$ & 1.086 &  \\ \hline
        PN A66 43 & 4488953930631143168 & 268.38446 & +10.62340 & 14.66 & $0.458 \pm 0.033$ & 7 & $2.09_{-0.11}^{+0.12}$ & $0.1946 \pm 0.0087$ & 1.038 & hybrid \\ \hline
        PN A66 72 & 1761341417799128320 & 312.50856 & +13.55817 & 16.01 & $0.548 \pm 0.064$ & 12 & $1.84_{-0.21}^{+0.18}$ & $0.06740 \pm 0.00065$ & 1.042 &  \\ \hline
        PN IsWe 1 & 250358801943821952 & 057.27473 & +50.00410 & 16.47 & $2.350 \pm 0.057$ & 2 & $0.424_{-0.009}^{+0.010}$ & $0.197 \pm 0.045$ & 0.903 &  \\ \hline
        PN Jn 1 & 2871119705335735552 & 353.97219 & +30.46843 & 16.00 & $1.011 \pm 0.065$ & 6 & $0.982_{-0.059}^{+0.071}$ & $0.0900 \pm 0.0044$ & 1.120 &  \\ \hline
        PN K 1$-$16 & 2160562927224840576 & 275.46708 & +64.36482 & 14.98 & $0.589 \pm 0.035$ & 6 & $1.737_{-0.092}^{+0.090}$ & $0.0388 \pm 0.0035$ & 1.102 &  \\ \hline
        PN Kn 12 & 1823929193070538624	& 300.84391 & +21.59786 & 18.44 & $0.33 \pm 0.17$ & \textbf{49} & $3.5_{-1.1}^{+1.5}$ & $0.2839 \pm 0.0071$ & 0.961	&  \\ \hline
        PN Kn 61 & 2052811676760671872 & 290.41223 & +38.31588 & 18.25 & $0.14 \pm 0.11$ & \textbf{80} & $5.9_{-1.8}^{+2.4}$ & $0.1327 \pm 0.0056$ & 0.986 & binary? \\ \hline
        PN Kn 130 & 1941078175572093696	& 348.27200 & +45.43838 & 16.54 & $0.497 \pm 0.054$ & 11 & $2.115_{-0.27}^{+0.27}$ & $0.1826 \pm 0.0022$ & 1.051 &  \\ \hline      
        PN Lo 3 & 5509004952576699904 & 108.70594 & $-46.96087$ & 16.74 & $0.467 \pm 0.074$ & 16 & $2.10_{-0.27}^{+0.30}$ & 0.172\tablenotemark{a} & \textbf{1.787} &  \\ \hline
        PN Lo 4 & 5414927915911816704 & 151.44074 & $-44.35931$ & 16.59 & $0.330 \pm 0.052$ & 16 & $3.06_{-0.40}^{+0.60}$ & 0.147\tablenotemark{a} & 1.039 &  \\ \hline
        PN Ou 2 & 430204780732841600 & 007.73643 & +61.40952 & 19.27 &	$0.77 \pm 0.22$ & \textbf{28} & $1.59_{-0.38}^{+0.82}$ & $0.3860 \pm 0.0072$ & 1.009 &  \\ \hline
        PN VV 47 & 936605992140011392 & 119.46507 & +53.42137 & 17.06 & $1.065 \pm 0.079$ & 7 & $0.985_{-0.076}^{+0.076}$ & $0.0371 \pm 0.0050$ & 1.008 &  \\ \hline
        RL 104 & 180006683580428928 & 067.56196 & +40.40398 & 13.71 & $0.947 \pm 0.021$ & 2 & $1.020_{-0.020}^{+0.025}$ & $0.3071 \pm 0.0035$ & 0.964 &  \\ \hline
        RX J0122.9$-$7521 & 4637921057358156416 & 020.72372 & $-75.35420$ & 15.38 & $1.196 \pm 0.035$ & 3 & $0.830_{-0.024}^{+0.023}$ & 0.053\tablenotemark{a} & 1.089 &  \\ \hline 
        RX J2117.1+3412 & 1855295171732158080 & 319.28448 & +34.20766 & 13.02 & $1.991 \pm 0.035$ & 2 & $0.4986_{-0.0094}^{+0.0082}$ & $0.0600 \pm 0.0021$ & 0.948 &  \\ \hline
        SALT J172411.7$-$632147 & 5910236846008692352 & 261.04877 & $-63.36322$ & 16.59 & $0.585 \pm 0.063$ & 11 & $1.78_{-0.19}^{+0.18}$ & 0.065\tablenotemark{a} & 0.936 &  \\ \hline
        SALT J213742.6$-$382901 & 6585736932806500736 & 324.42712 & $-38.48355$ & 16.95 & $0.538 \pm 0.087$ & 16 & $1.94_{-0.26}^{+0.36}$ & 0.036\tablenotemark{a} & 1.061 &  \\ \hline
        SDSS J000945.46+135814.4 & 2767982864653184640 & 002.43941 & +13.97065 & 18.07 & $0.31 \pm 0.17$ & \textbf{55} & $2.6_{-0.7}^{+1.3}$ & $0.0829 \pm 0.0041$ & 0.989 &  \\ \hline
        SDSS J001651.42$-$011329.3 & 2541718902258404736 & 004.21425 & $-01.22487$ & 16.75 & $0.273 \pm 0.079$ & \textbf{29} & $3.36_{-0.60}^{+0.81}$ & $0.06729 \pm 0.00039$ & 0.995 &  \\ \hline
        SDSS J034917.41$-$005919.3 & 3251245339191040256 & 057.32256 & $-00.98874$ & 17.80 & $1.15 \pm 0.12$ & 10 & $0.85_{-0.08}^{+0.12}$ & $0.1274 \pm 0.0061$ & 1.048 &  \\ \hline
        SDSS J055905.02+633448.4 & 286746241613044096 & 089.77088 & +63.58012 & 18.59 & $0.98 \pm 0.16$ & 16 & $1.06_{-0.14}^{+0.17}$ & $0.1606 \pm 0.0025$ & 1.078 &  \\ \hline
        SDSS J075415.11+085232.1 & 3145662944130394496 & 118.56299 & +08.87560 & 19.08 & $0.57 \pm 0.23$ & \textbf{39} & $1.8_{-0.5}^{+1.2}$ & $0.0321 \pm 0.0065$ & 0.961 &  \\ \hline
        SDSS J075540.94+400918.0 & 920621124593362816 & 118.92053 & +40.15497 & 17.80 & $0.95 \pm 0.13$ & 13 & $1.14_{-0.13}^{+0.19}$ & $0.0572 \pm 0.0041$ & 0.968 &  \\ \hline
        SDSS J093546.53+110529.0 & 589674614326779136 & 143.94384 & +11.09133 & 17.75 & $1.10 \pm 0.14$ & 13 & $0.96_{-0.10}^{+0.13}$ & $0.0411 \pm 0.0030$ & 0.985 &  \\ \hline
        SDSS J102327.41+535258.7 & 851812381256776832 & 155.86423 & +53.88297 & 17.92 & $0.50 \pm 0.11$ & \textbf{23} & $2.03_{-0.37}^{+0.47}$ & $0.0280 \pm 0.0028$ & 0.977 &  \\ \hline
        SDSS J105300.24+174932.9 & 3982986781494206080 & 163.25103 & +17.82578 & 16.76 & $1.429 \pm 0.076$ & 5 & $0.714_{-0.034}^{+0.040}$ & $0.0190 \pm 0.0016$ & 0.987 &  \\ \hline
        SDSS J121523.09+120300.8 & 3908341899157118080 & 183.84614 & +12.05020 & 18.14 & $0.75 \pm 0.14$ & 19 & $1.44_{-0.23}^{+0.35}$ & $0.03492 \pm 0.00078$ & 0.928 &  \\ \hline
        SDSS J123930.61+244321.7 & 3959650269965155584 & 189.87752 & +24.72270 & 18.30 & $0.69 \pm 0.16$ & \textbf{23} & $1.51_{-0.26}^{+0.37}$ & $0.0405 \pm 0.0090$ & 0.975 &  \\ \hline
        SDSS J134341.88+670154.5 & 1672427588951276800 & 205.92436 & +67.03180 & 17.13 & $1.455 \pm 0.055$  & 4 & $0.707_{-0.026}^{+0.023}$ & $0.0284 \pm 0.0014$ & 0.937 &  \\ \hline
        SDSS J141556.26+061822.5 & 3673120627847661184 & 213.98441 & +06.30622 & 17.44 & $1.04 \pm 0.13$ & 12 & $1.00_{-0.11}^{+0.14}$ & $0.0303 \pm 0.0055$ & 1.102 &  \\ \hline
        SDSS J144734.12+572053.1 & 1613731019696686208 & 221.89206 & +57.34807 & 18.03 & $0.835 \pm 0.092$ & 11 & $1.23_{-0.13}^{+0.21}$ & $0.0342 \pm 0.0082$ & 1.028 &  \\ \hline
        SDSS J152116.00+251437.5 & 1270099761612163328 & 230.31665 & +25.24375 & 17.87 & $0.26 \pm 0.11$ & \textbf{43} & $4.4_{-1.4}^{+2.1}$ & $0.0386 \pm 0.0066$ & 1.058 & hybrid \\ \hline
        SDSS J155610.40+254640.3 & 1220049614357436544 & 239.04334 & +25.77784 & 17.91 & $0.086 \pm 0.098$ & \textbf{115} & $7.5_{-3.0}^{+3.3}$ & $0.0630 \pm 0.0077$ & 0.981 & binary? \\ \hline
        SDSS J163727.03+485355.2 & 1410694377877399552 & 249.36262 & +48.89866 & 18.35 & $0.57 \pm 0.11$ & 19 & $2.01_{-0.36}^{+0.84}$ & $0.0261 \pm 0.0055$ & 1.018 &  \\ \hline
        SDSS J191845.01+624343.7 & 2240494910007892608 & 289.68757 & +62.72883 & 17.58 & $0.970 \pm 0.074$ & 8 & $1.111_{-0.079}^{+0.101}$ & $0.0262 \pm 0.0020$ & 1.012 &  \\ \hline
        SDSS J212531.92$-$010745.8 & 2686081102494206080 & 321.38303 & $-01.12941$ & 17.54 & $0.35 \pm 0.11$ & \textbf{32} & $2.9_{-0.7}^{+1.3}$ & $0.03730 \pm 0.00084$ & 1.034 & binary \\ \hline
        Sh 2$-$68 & 4276328581046447104 & 276.24337 & +00.85976 & 16.40 & $2.446 \pm 0.059$ & 2 & $0.405_{-0.010}^{+0.010}$ & $0.622 \pm 0.048$ & 1.088 & hybrid \\ \hline
        Sh 2$-$78 & 4506484097383382272 & 285.79198 & +14.11631 & 17.61 & $1.43 \pm 0.10$ & 7 & $0.696_{-0.044}^{+0.059}$ & $0.3160 \pm 0.0085$ & 1.045 &  \\ \hline
        TIC 95332541 & 2997192526074656640 & 090.68749 & $-13.85096$ & 15.32 & $2.593 \pm 0.043$ & 2 & $0.3845_{-0.0050}^{+0.0055}$ & $0.0575 \pm 0.0049$ & 1.023 &  \\ \hline
        TIC 333432673 & 2950907725113997312 & 100.31517 & $-13.69000$ & 15.21 & $2.552 \pm 0.043$ & 2 & $0.3892_{-0.0054}^{+0.0054}$ & $0.119 \pm 0.018$ & 1.093 &  \\ \hline 
        TIC 403800675 & 3486203758501245440 & 179.36518 & $-28.06384$ & 16.16 & $1.875 \pm 0.062$ & 3 & $0.535_{-0.018}^{+0.019}$ & $0.0591 \pm 0.0040$ & 1.004 &  \\ \hline
        TIC 1989122424 & 6462935326662402944 & 319.40996 & $-55.46694$ & 16.75 & $1.471 \pm 0.062$ & 4 & $0.688_{-0.026}^{+0.022}$ & 0.058\tablenotemark{a} & 0.987 &  \\ \hline
        WD J070204.29+051420.56 & 3128765207057429504
 & 105.51783 & +5.23904 & 14.98 & $3.089 \pm 0.053$ & 2 & $0.3228_{-0.0057}^{+0.0053}$ & $0.0472 \pm 0.0015$ & 1.091 & \\ \hline \hline
        NGC 6765 & 2039515046435901440 & 287.77732 & +30.54545 & 17.60 & $0.276 \pm 0.078$ & \textbf{28} & $4.0_{-1.0}^{+1.5}$ & $0.1505 \pm 0.0023$ & 1.002 &  \\ \hline
        PG 2131+066 & - & - & - & - & - & - & - & - & - & binary \\ \hline
        RX J0439.8$-$6809 & - & - & - & - & - & - & - & - & - & 
\enddata
\tablenotetext{a}{Reddening from SFD}
\tablecomments{Astrometric properties of PG 1159 stars from \textit{Gaia} DR3 \citep{2023AA...674A...1G}. Geometric distances (rgeo) are from \citet{2021AJ....161..147B}, E(B-V) from Bayestar17 \citep{2018MNRAS.478..651G} except for targets with Dec. south of $-30^{\circ}$ where SFD maps \citep{2011ApJ...737..103S,1998ApJ...500..525S} were used. Uncertainties in E(B-V) are calculated as half of the difference between values at 16th and 84th percentile. The last three objects were excluded from the analysis because of the lack of either \textit{Gaia} measurements, or $T_{\mathrm{eff}}$ and $\log{g}$.}
\end{deluxetable*}
\end{longrotatetable}


\section{PG~1159 stars on the Hertzsprung-Russell diagram}
\label{sec:hrd}

\begin{figure}
	\includegraphics[width=\linewidth]{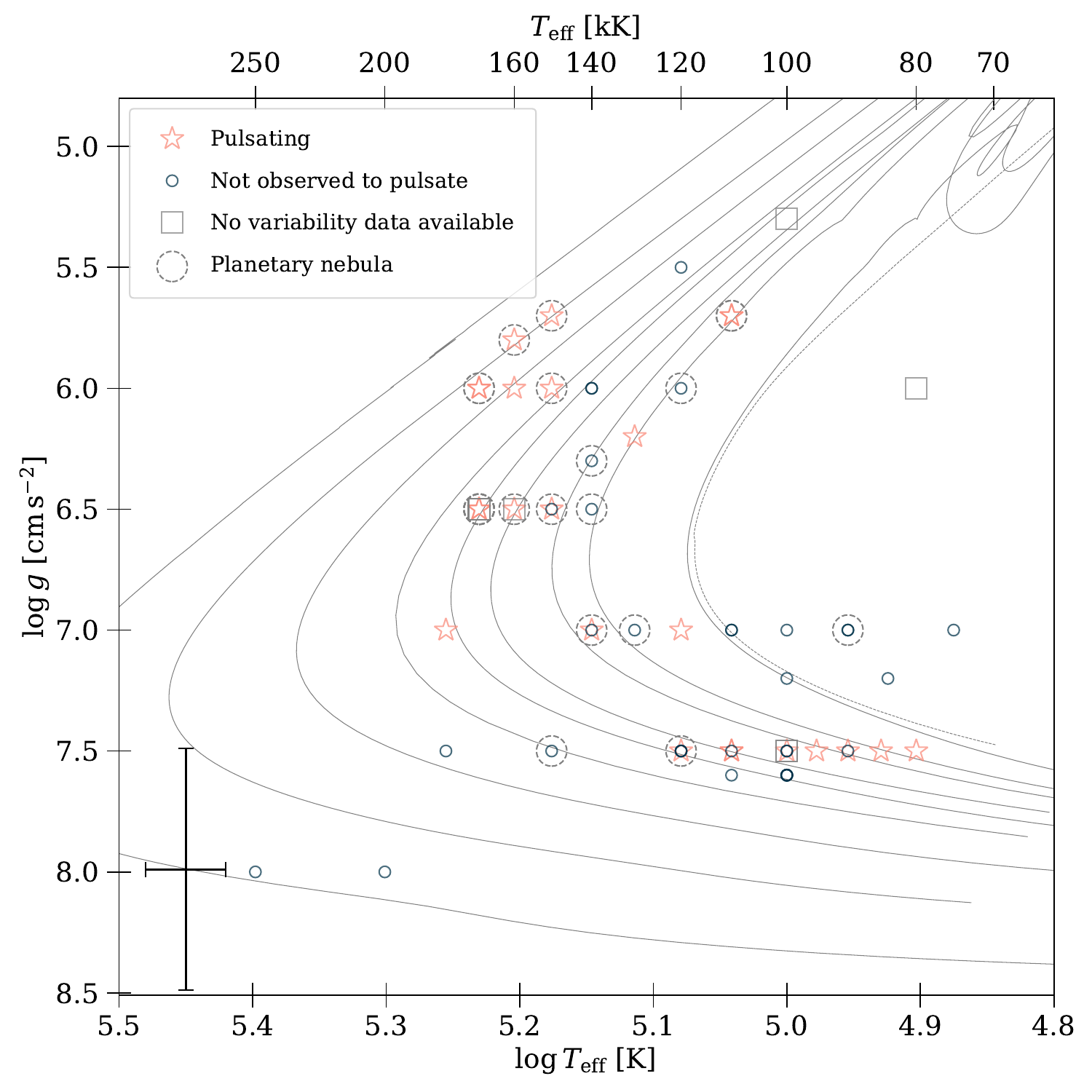}
    \caption{Positions of PG~1159 stars in the surface gravity-effective temperature $\log{g}-\log{}T_{\mathrm{eff}}$ diagram. Star symbols represent pulsating PG~1159 stars, circles – non-variable, and squares – with no reported photometric observations. Stars with planetary nebulae are marked with dashed circles. A typical error bar is shown in the bottom left corner. Multiple stars overlap in this diagram, which can be seen as wider and darker borders of symbols. See text for details. Lines represent evolutionary tracks from \citet{2006AA...454..845M}: solid -- VLTP (from left to right, final masses: 0.870, 0.741, 0.664, 0.609, 0.584, 0.565, 0.542, 0.530, 0.515 M$_{\odot}$), single dashed line -- LTP (0.512 M$_{\odot}$).}
    \label{fig:PG1159_Kiel}
\end{figure}

PG 1159 stars plotted in the surface gravity-effective temperature diagram $\log{g}-\log{}T_{\mathrm{eff}}$ (also called the Kiel diagram) cluster horizontally along the lines of constant $\log{g}$, and vertically along the lines of constant $\log{}T_{\mathrm{eff}}$ (see Figure~\ref{fig:PG1159_Kiel}). The reasons are the current sensitivity of spectroscopic observations (large uncertainties, for some PG~1159 stars even $\pm0.5$~\cms in $\log{g}$) and availability of advanced model atmospheres for these extremely hot stars, with the latter usually provided in grids with a step of $\log{g} = 0.5$ \cms and $T_{\mathrm{eff}}=10000~\mathrm{K}$.  

For the discussion in the context of asteroseismology it is useful to place the PG 1159 stars into the theoretical Hertzsprung-Russell diagram ($\log{}L_{\star} / L_{\odot}-\log{}T_{\mathrm{eff}}$). However, this requires the knowledge of stellar luminosities and effective temperatures. 
Derivation of stellar luminosities is especially challenging, because it relies on knowing the total bolometric flux of a given star. There are many ways to tackle this difficult problem. 
One solution is based on determining the Spectral Energy Distribution (SED) by fitting model atmospheres to broadband photometric magnitudes (see, e.g., \citealt{2022MNRAS.513.2285U}). For such hot stars as PG~1159s, the UV photometry and a grid of model atmospheres covering those short wavelengths is essential and, to date, not available for the whole sample of PG~1159 stars. 
 
Another method is to derive the bolometric luminosities, either from mathematical prescriptions, or from apparent magnitudes using bolometric corrections (BCs). 
In the second case, the observed apparent magnitudes are converted to absolute magnitudes in a given passband $b$ using a distance modulus $DM$ (a logarithmic measure of the distance to the star):
\begin{equation}
\label{eq:first}
    m_b = M_b + DM,
\end{equation}
where $m_b$ is the apparent magnitude and $M_b$ is the absolute magnitude in the passband $b$. Incorporating the definition of the absolute magnitude gives:
\begin{equation}
\label{eq:dm}
    DM = m_b - M_b = 5 \log{}_{10} \frac{d}{\mathrm{(10 pc)}},
\end{equation}
where $d$ is distance in parsecs. With the correction for interstellar absorption between the object and observer, the absolute magnitude in a passband $b$ can be derived from: 
\begin{equation}
    M_b = m_b - DM - A_b, 
\end{equation}
where $A_b$ is the extinction in a passband $b$. 
Then the bolometric magnitude is
\begin{equation}
    M_{\mathrm{bol}} = M_b + BC_b,
\end{equation}
where $M_{\mathrm{bol}}$ is absolute bolometric magnitude and $BC_b$ is bolometric correction in a given passband, a quantity dependent not only on the photometric passband used in observations, but also the theoretical stellar spectrum used in calculation of the correction (different sets of effective temperature, surface gravity, and metallicity will give different BC values). In the case of extremely hot stars such as pre-white dwarfs of the PG 1159 type, this requires using models including non-local thermodynamic equilibrium (non-LTE) effects. 
Finally, the absolute bolometric magnitude $M_{\mathrm{bol}}$ of a star of a bolometric luminosity $L_{\star}$, referenced to the Sun, is given by:
\begin{equation}
\label{eq:last}
    -2.5 \log_{10} \frac{L_{\star}}{L_{\odot}} = M_{\mathrm{bol}} - M_{\mathrm{bol},\odot}, 
\end{equation}
where $M_{\mathrm{bol},\odot} = 4.74$ is the absolute bolometric magnitude of the Sun, and $L_{\odot} = 3.828 \times 10^{33}\, \mathrm{erg}\,  \mathrm{s}^{-1}$ is the absolute bolometric luminosity of the Sun\footnote{IAU Resolution 2015 B2}. 

For white dwarf stars, the first commonly used/tabulated BC values were compiled by \citet{1995PASP..107.1047B} for hydrogen- and helium-rich white dwarf model atmospheres, but for a small grid covering surface gravity (only $\log{g}=8$~\cms) and effective temperature (up to $100000\, \mathrm{K}$ only for DA white dwarfs). This work was expanded by \citet{2006AJ....132.1221H}, who provided an extensive grid for both DA and DB white dwarfs\footnote{The DA grid covered $T_{\mathrm{eff}} = 2500$ K to 150000 K and $\log{g}$ = 7.0 to 9.0 \cms, while the DB grid covered $T_{\mathrm{eff}} = 3250$ K to 150000 K and $\log{g}$ = 7.0 to 9.0 \cms.}. The latter work is regularly updated on line\footnote{\url{https://www.astro.umontreal.ca/\~bergeron/CoolingModels/} }. In the most up-to-date version of the tables, models of \citet{2020ApJ...901...93B}, which include non-LTE effects, are used at the highest effective temperatures. Unfortunately, no bolometric corrections have ever been extensively compiled for PG 1159 stars. Some PG 1159 stars had bolometric corrections estimated for the purpose of deriving luminosities for asteroseismic modeling (\citealt{2021AA...655A..27U} list three previously used values), but no tabulated prescription has ever been provided. 

We calculated the luminosities of PG~1159 stars based on currently available data. We used the distances and interstellar reddening values described in Section~\ref{s:PG1159properties}. The reddening for each star was converted to extinction using the reddening law of \citet{2004ASPC..309...33F} with $R_V = 3.1$. \textit{Gaia} magnitudes were converted to $V$ using the following prescription\footnote{\href{https://gea.esac.esa.int/archive/documentation/GDR3/Data_processing/chap_cu5pho/cu5pho_sec_photSystem/cu5pho_ssec_photRelations.html\#Ch5.T9}{Gaia DR3 documentation.} We note that a few objects were slightly outside the range of applicability for this relationship.}:
\begin{eqnarray}
G - V &= -0.02704 + 0.01424(G_{\mathrm{BP}}-G_{\mathrm{RP}}) \\
&- 0.2156(G_{\mathrm{BP}}-G_{\mathrm{RP}})^2 \nonumber \\ 
&+ 0.01426(G_{\mathrm{BP}}-G_{\mathrm{RP}})^3.  \nonumber    
\end{eqnarray}
We used tabulated bolometric corrections for pure-helium model atmospheres (DB) provided online on the aforementioned website by the Montreal group. As the bolometric correction primarily depends on the effective temperature and because there are no bolometric corrections computed with proper models for PG~1159 stars, we used those models as the best approach currently available. 
The tabulated values do not cover surface gravities below $\log{g}=7.0$~\cms, therefore the ones for $\log{g}=7.0$~\cms were used for matching effective temperatures. The values for effective temperatures over 150000~K were extrapolated to higher effective temperatures for a given $\log{g}$. The linear extrapolation was done in $\log{T_{\mathrm{eff}}}$ vs. BC space using \texttt{interp1d} class from the \texttt{scipy} sub-package \texttt{interpolate} and ``fill\_value=`extrapolate'\,'', using the available BC values for $T_{\mathrm{eff}}$ in the range $75000 - 150000$ K for a given $\log{g}$. Table~\ref{tab:PG1159_physical} lists the physical properties and chosen BC values for each PG~1159 star in the sample. Then, the luminosities were calculated following equations \ref{eq:first} -- \ref{eq:last} and are also listed in Table~\ref{tab:PG1159_physical} with uncertainties. The errors were propagated the following way: a) for $DM$ using asymmetric errors from Table~\ref{tab:astrometry}, b) for E(B-V) using symmetric errors from Table~\ref{tab:astrometry}, c) for $BC_V$ using asymmetric errors adopted as the $BC_V$ values $\pm10000$~K for each object, d) for $G$, $G_{\mathrm{BP}}$, and $G_{\mathrm{RP}}$ magnitudes the symmetric errors were calculated as $1.09 \cdot G / \mathrm{SNR}$, where SNR is roughly \texttt{phot\_g\_mean\_flux\_over\_error}\footnote{\url{https://dc.zah.uni-heidelberg.de/gaia/q3/cone/info\#note-e}} (example for $G$).

\begin{figure*}
	\includegraphics[width=\textwidth]{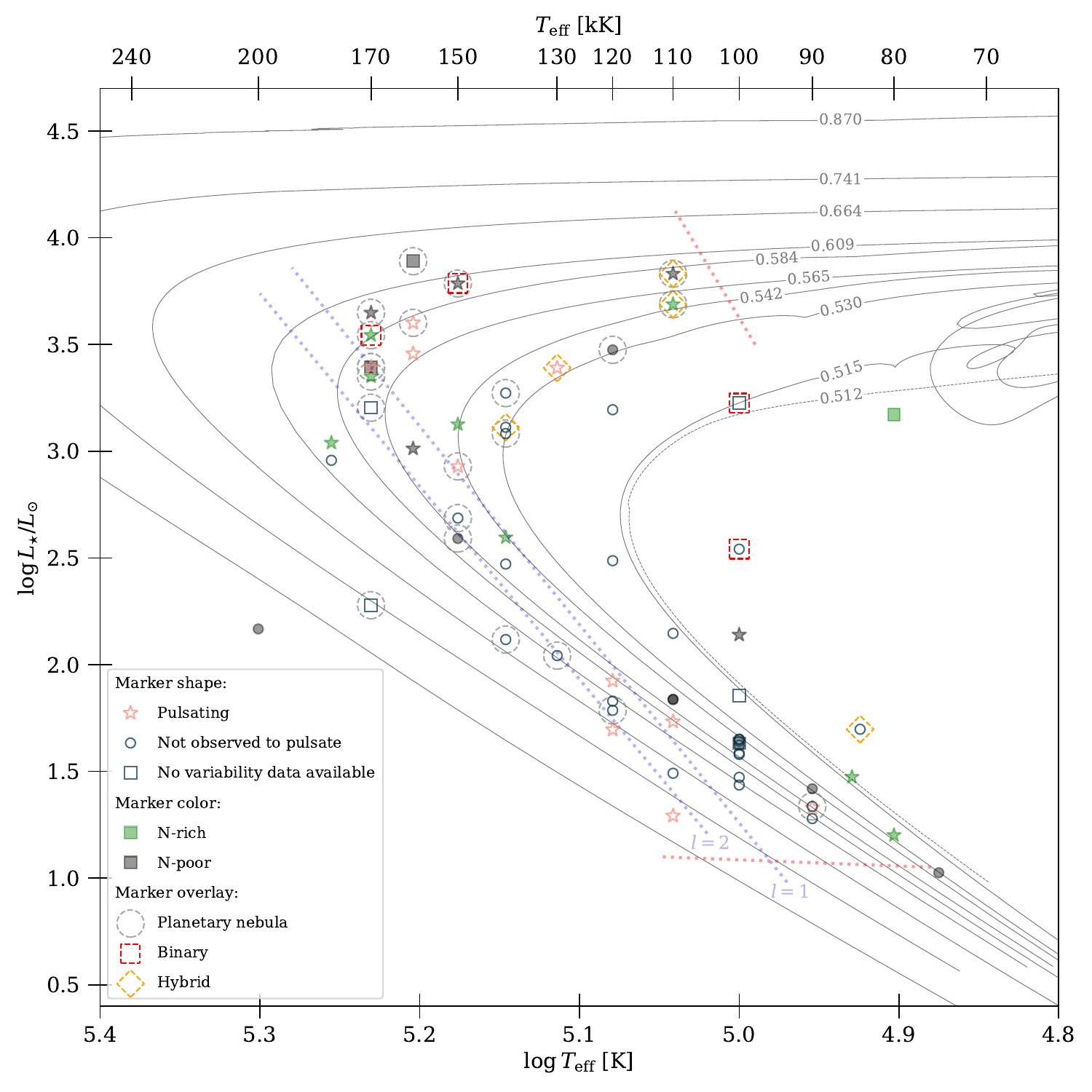}
    \caption{Positions of PG~1159 stars in the theoretical Hertzsprung-Russell diagram ($\log{}L_{\star} / L_{\odot}-\log{}T_{\mathrm{eff}}$). Star symbols represent pulsating PG~1159 stars, circles – non-variable, and squares – with no reported photometric observations. Stars with planetary nebulae are marked with dashed circles. N-rich PG~1159 stars are shown with filled green symbols, while N-poor ones with filled black symbols. Lines represent evolutionary tracks from \citet{2006AA...454..845M}: solid -- VLTP (from left to right, final masses: 0.870, 0.741, 0.664, 0.609, 0.584, 0.565, 0.542, 0.530, 0.515 M$_{\odot}$), single dashed line -- LTP (0.512 M$_{\odot}$). For illustration purposes, the blue dotted lines represent theoretical blue edges for $l=1$ and $l=2$ modes from \citet{2005AA...438.1013G}, but the blue edge is composition dependent. The red dotted lines indicate estimated observed red edges, beyond which no GW Vir star has to date been reported.}
    \label{fig:PG1159_HR}
\end{figure*}

Figure~\ref{fig:PG1159_HR} shows positions of PG~1159 stars in the theoretical Hertzsprung-Russell diagram ($\log{}L_{\star} / L_{\odot}-\log{}T_{\mathrm{eff}}$). For illustration purposes, the blue dotted lines represent theoretical blue edges for $l=1$ and $l=2$ modes from \citet{2005AA...438.1013G}, but the blue edge is composition dependent – ``fuzzy'' \citep{2007ApJS..171..219Q}, and with the red dotted lines we show the presently observed red edges.

\section{Discussion}

The number of pulsating PG~1159 stars increased to 24 objects with our discovery of pulsations in Abell 72. The main observational challenge in the detection or confirmation of variability in those stars lies in two main areas. Firstly, the amplitudes of the g-mode pulsations are quite low. While Abell 72 showed pulsation amplitudes of up to 10~mmag, PG~1144+005, on the other hand, showed a highly variable (between consecutive nights) Fourier spectrum with amplitudes ranging from 3  to 6 mmag \citep{2021ApJ...918L...1S}. This requires reaching a noise level below 1 mmag for a significant detection (assuming S/N$\geq$4), which is a challenging task for these faint stars. We were not able to reach noise levels below 1.5 mmag for Longmore 3, SDSS J000945.46+135814.4, SDSS J001651.42$-$011329.3, SDSS J102327.41+535258.7, and SDSS J144734.12+572053.1. Another challenge is the aforementioned change in amplitude spectra for some stars, between observing seasons or even consecutive nights. It is therefore always possible that the star is observed in a temporary state where the pulsations destructively interfere. We aimed at obtaining more than one run for each star in the sample with a sufficient quality, but this was only possible for eight stars. 

Our results allowed us to update the fraction of PG~1159 pulsators in the GW Vir instability strip. While previous works quoted values of about $20-50\%$, but including not only PG~1159 stars but also the other stars populating the GW Vir instability strip, we obtain 36\% using only stars of PG 1159 spectral type. Our fraction is consistent with previous estimates and shows that only about 1/3 of PG~1159 stars within the GW Vir instability strip are observed to vary.

In this context, it is interesting to see how the variability compares to the nitrogen abundance observed in PG~1159 stars, in the light of the nitrogen dichotomy (N-rich pulsators, N-poor nonpulsators, found by \citealt{1998AA...334..618D}) that appears to hold. While the majority of those stars do not have a determination of their atmospheric nitrogen abundance available in the literature, there are a few stars that may not fit this hypothesis. The most recent analysis of the pulsating central star of NGC~246 by \citet{2018Galax...6...65L} implies sub-solar N abundance. 
SALT~J172411.7$-$632147 is a N-poor pulsator reported by \citet{2023MNRAS.519.2321J}. New spectra of TIC~95332541 analyzed by \citet{2023arXiv230703721R} revealed that it is another N-poor pulsator. 
Longmore 4 is a known pulsator, and does not show N in a number of medium-resolution spectra. It is interesting in the context of the outbursts that it exhibits, temporarily changing its spectral type from PG~1159 to [WCE] \citep{1992AA...259L..69W,2014AJ....148...44B}. 
RL~104 is also an interesting object, as it is N-rich and claimed to have evolved from a binary merger scenario, but to date was not observed photometrically. 
 
With such a sample tested for variability, we placed the PG~1159 stars in the theoretical Hertzsprung-Russell diagram. We determined luminosities following the procedure described in Sect.~\ref{sec:hrd}. We plotted them against available evolutionary tracks for PG~1159 stars. In general, very good agreement between the evolutionary tracks and positions of PG~1159 stars was obtained. The majority of the stars are within the evolutionary tracks for typical PG~1159 masses ($0.5-0.6$~M$_{\odot}$). Only one star is found beyond 0.87~M$_{\odot}$ -- H1504+65. Nevertheless, a few shortcomings of our attempt need to be noted. The distances from \textit{Gaia} for some stars have large uncertainties due to large relative errors of parallaxes. In Table~\ref{tab:astrometry} we marked 14 stars whose relative parallax errors exceed 20\%. Four of them are confirmed or suspected binaries, therefore their determined positions might be uncertain. It is worth comparing the distances determined using different (independent) methods, e.g., using planetary nebulae line strengths, but this is out of the scope of this work. 
\citet{2021AA...655A..27U} quoted available in the literature values of bolometric correction for PG~1159 stars for three objects: PG~1159$-$035 (T$_{\mathrm{eff}}=140000$~K, $\log{g}=7.0$~\cms): $BC=-7.6$, RX~J2117+3142 (T$_{\mathrm{eff}}=170000$~K, $\log{g}=6.0$~\cms): $BC=-7.95$, and PG~2131+066 (T$_{\mathrm{eff}}=95000$~K, $\log{g}=7.5$~\cms): $BC=-6.0$. They interpolated those values to obtain $BC=-7.05$ for TIC 95332541 and TIC 333432673 (T$_{\mathrm{eff}}=120000$, $\log{g}=7.5$), assuming only the dependence on the effective temperature. We investigated the difference between tabulated BC for DA and DB models. For T$_{\mathrm{eff}}$ and $\log{g}$ of PG~1159$-$035 (the only exact match with tabulated values), we found BC of $-7.964$ and $-7.658$ for DA and DB models, respectively. The value for the DB model atmosphere agrees well with the quoted value of $BC=-7.6$. We also checked how the BC value from DB table changes with $\log{g}$ for a given temperature. For T$_{\mathrm{eff}}=140000$~K and $\log{g}=7.0, 7.5, 8.0, 8.5$~\cms we obtained $BC=-7.658, -7.668, -7.676, -7.681$, respectively. Therefore, we do not expect significant interpolation errors in the parameter space of interest.


\section{One or two GW Vir instability domains?}

The establishment of the PG 1159 spectral class (e.g., see \citealt{1992LNP...401..273W}) occurred subsequently to the discovery of pulsations in PG 1159$-$035 \citep{1979wdvd.coll..377M} itself. At the time when the pulsating PG 1159 stars emerged as a new group of pulsators (e.g., \citealt{1984ApJ...279..751B}), they were considered the hottest subgroup of the helium-rich DO white dwarf stars \citep{1985ApJS...58..379W}. For that reason, and for the similarity with the designations of the groups of pulsating white dwarfs already known (DAV and DBV) the PG 1159 pulsators were dubbed ``the DOVs''.

However, the second pulsating star of the PG 1159 spectral type discovered was located in a planetary nebula \citep{1984ApJ...277..211G} and subsequent searches (e.g., \citealt{1996AJ....111.2332C}) revealed several of these ``Planetary Nebula Nucleus Variables'' (PNNVs). Even though it was realized that the ``DOVs'' were likely just the same type of pulsating stars, but in a more advanced evolutionary stage than the ``PNNVs'', the two groups were historically often separated. The main reason for this separation was that one group of pulsators are surrounded by nebulae whereas the others were not and that one group has significantly longer pulsation periods than the other. Furthermore, theoretical computations (e.g., \citealt{2006AA...458..259C}, cf. Fig.\,\ref{fig:PG1159_HR}) show that the blue edge of the instability strip intersects with the evolutionary tracks of pre-white dwarf stars in such a way that many of them leave the strip during their evolution and later re-enter it, giving the impression of two separated instability regions.

\cite{2007ApJS..171..219Q} and \cite{2008PASP..120.1043F} argued, mostly on a theoretical basis, that this separation should be dropped, and that all hot pulsating pre-white dwarf stars should be called ``the GW Vir stars''\footnote{GW Vir is the variable star designation of PG 1159$-$035, \cite{1985IBVS.2681....1K}.}. This was motivated by the fact that the pulsational driving mechanism of all GW Vir stars is the same, that stars with a pure DO spectral type are not known to pulsate\footnote{The PG 1159 spectral class had meanwhile been established as a separate group, and we recall that some pulsating pre-white dwarfs are of [WCE] or [WCE/PG 1159] spectral types.}, and that not all stars classified as ``PNNV'' even possess a detected planetary nebula. To this it can be added that there are other intrinsically variable central stars of planetary nebulae (e.g., \citealt{2013MNRAS.430.2923H}) that do not pulsate at all, which is why a designation ``PNNV'' is equally misleading as is ``DOV''.

\citet{1990AA...231L..33S} showed that some PNNs are spectroscopically indistinguishable from the white dwarfs similar to PG 1159$-$035, and assigned them all to ``PG 1159'' type. PG~1159 subclasses introduced by \citet{1992LNP...401..273W} did not take into account the presence or absence of a nebula, treating the PG~1159 spectral class as a whole. Therefore, not only is the pulsation driving mechanism the same for those stars, but they also share some spectroscopic properties representative of the whole class. 

Moreover, the commonly used surface gravity-effective temperature diagram presented in Fig.~\ref{fig:PG1159_Kiel} shows that it is impossible to separate the two groups in the $\log{g}-\log{}T_{\mathrm{eff}}$ plane – stars with planetary nebulae are found throughout the whole GW Vir instability strip. This refutes the argument that the PNNVs usually have much lower surface gravities, as no strict boundary can be placed in such a plane.

\begin{figure*}
    \centering
    \includegraphics[width=0.9\textwidth]{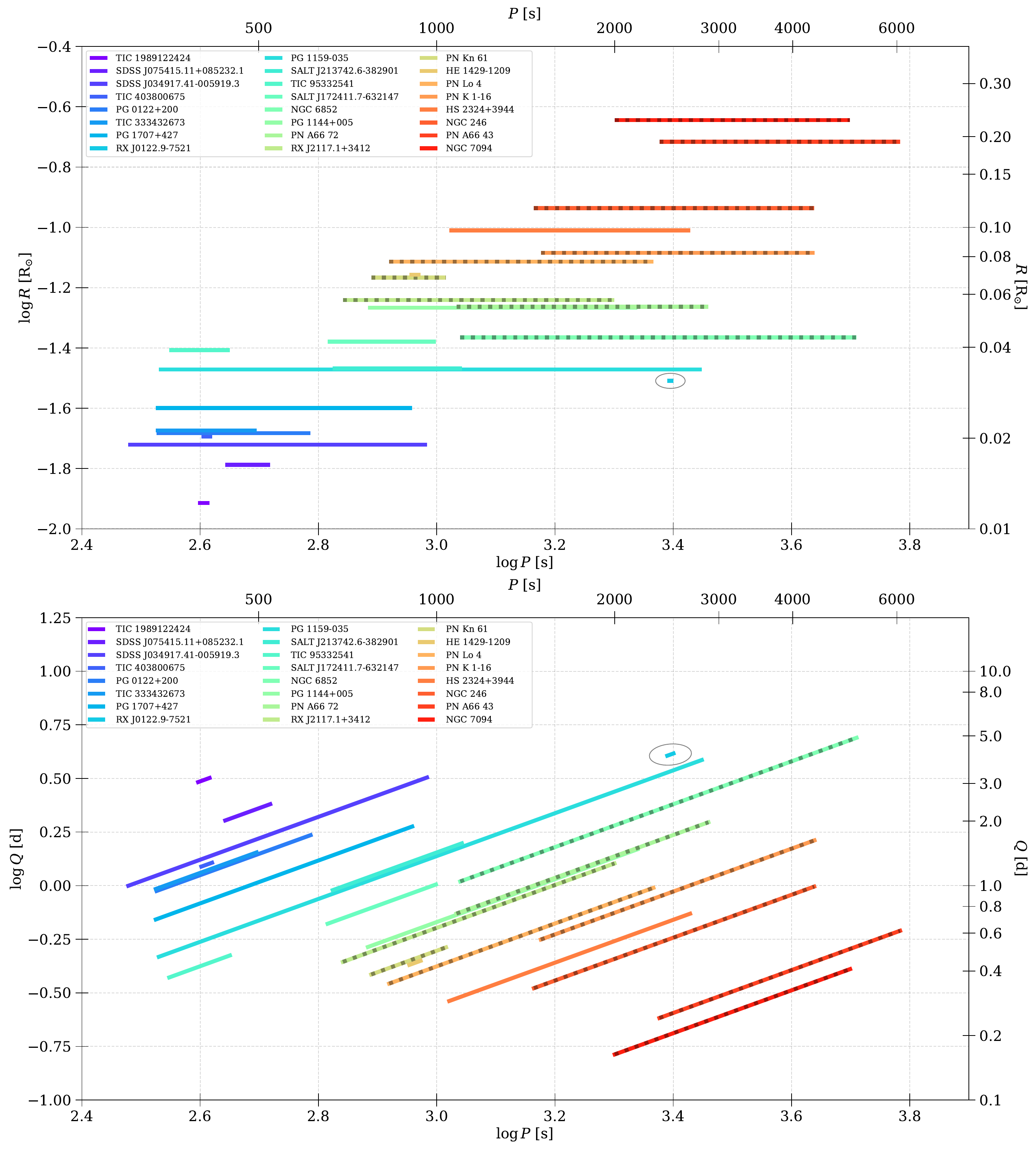}
    \caption{\textit{Top:} The period ranges of pulsating pre-white dwarf stars (only of PG~1159 spectral type) versus stellar radius (horizontal bars). \textit{Bottom:} The period ranges of pulsating pre-white dwarf stars (only of PG~1159 spectral type) versus pulsation constant (horizontal bars). Objects surrounded by a planetary nebula are denoted with grey-dotted bars. The object marked with an ellipse is RX J0122.9$-$7521 (see Sect.\,\ref{rxj0122}).}
    \label{fig:period-radius-Q}
\end{figure*}

In the following, we examine the question whether these two groups are distinct, or should be distinguished, from an observational point of view. The top panel in Figure\,\ref{fig:period-radius-Q} shows the ranges of pulsation periods observed in pulsating pre-white dwarf stars (of PG 1159 spectral type) versus stellar radius (as derived from Fig.\,\ref{fig:PG1159_HR} and the Stefan-Boltzmann law).

Several things are noteworthy in Fig.\,\ref{fig:period-radius-Q}. First of all, there is a clear overlap between the objects with and without a planetary nebula, already suggesting these two groups are not distinct. Second, an obvious trend as already noticed by others earlier is visible, namely that the larger, less evolved objects have longer pulsation periods.

To look into this in some more detail, the bottom panel in Fig.\,\ref{fig:period-radius-Q} compares the pulsation periods with the pulsation constant $Q=P\sqrt{\rho_{\ast}/\rho_{\odot}}$, a measure of the radial overtone of the excited pulsation modes. There is a slight trend that the longer the pulsation period, the smaller is the pulsation constant. In other words, the more evolved a pulsating pre-white dwarf, the higher radial overtones of the gravity modes are excited. This is consistent with the theoretical expectation that with progressing evolution the pulsational driving region becomes located closer to the stellar surface (see \citealt{2005AA...438.1013G}).

We therefore conclude that the distinction between ``DOVs'' and ``PNNVs'' is, according to current knowledge, artificial and based on selection effects and hence should not be used. All pulsating pre-white dwarf stars oscillating in gravity modes excited by the $\kappa-\gamma$ mechanism due to ionization of carbon and oxygen should henceforth be called ``GW Vir stars''.

\subsection{The case of RX J0122.9$-$7521}\label{rxj0122}

In Sect.\,\ref{sec:cand} we reported the detection of variability of RX J0122.9$-$7521 and mentioned that it would be the hottest GW Vir pulsator. However, we are reluctant to claim the firm detection of pulsation for this star, for several reasons.

Although its 41-min period fits in the range of pulsation periods observed in GW Vir stars \citep{2010AARv..18..471A}, we detected only a single period that could therefore in principle be of a different origin, like rotation, binarity, or spots \citep{2021AA...647A.184R}.
Furthermore, in Fig.\,\ref{fig:period-radius-Q}, the period of this star is rather long with respect to stars with similar radii.

 RX J0122.9$-$7521 lies outside the theoretical blue edge of GW Vir instability strip, and \citet{2004ApJ...610..436Q} unsurprisingly did not find an asteroseismic model with unstable periods in this star. \citet{1995BaltA...4..340W} reported the detection of nitrogen in its spectrum.

In any case, time-resolved spectroscopy or high signal-to-noise photometry would be needed to establish the cause of the variability of RX J0122.9$-$7521.


\section{Summary and conclusions}

We obtained new photometric observations of 29 PG~1159 stars. Over 86 hours of time-series photometry were collected in the years $2014-2022$ using telescopes of different sizes, ranging from 1.0-m to 10.4-m, and located in both hemispheres. For the majority of stars we achieved a median noise level in Fourier amplitude spectra in the range $0.3-1.0$~mmag, which allowed us to discover multiperiodic pulsations in the central star of planetary nebula Abell 72, and variability in RX J0122.9$-$7521 that could be due to pulsations, binarity or rotation. Five stars showed interesting peaks but require follow-up observations for confirmation. For the remaining stars our observations put limits on nonvariability. As a result, we derived the fraction of pulsating PG~1159 stars within the GW Vir instability strip -- 36\%. 

In the light of N dichotomy in PG~1159 stars, we compared the new variability results with the literature data on N abundances for those stars and identified objects that could be culprits for this hypothesis: NGC 246, SALT J172411.7$-$632147 and TIC 95332541 may be N-poor pulsators. Longmore 4 is probably a N-poor pulsator, but temporarily changes its spectral type from PG~1159 to [WCE] during outbursts.

Taking advantage of the currently available data, we used distances derived from $Gaia$ parallaxes, interstellar extinction from 3D reddening maps, and bolometric correction values from DB tables, to derive luminosities and place the PG~1159 stars in the theoretical Hertzsprung-Russell diagram. Regardless of the possible caveats of our approach, all stars align well with the PG~1159 evolutionary tracks from \citet{2006AA...454..845M}. 

Finally, we derived radii and pulsation constants for known pulsators, and plotted them against period ranges observed in those stars to further argue against the distinction between ``DOVs'' and ``PNNVs'', and suggested using only the ``GW Vir'' designation for all stars belonging to that family of pulsating white dwarfs.


\section{Author contributions}

PS with GH and DJ applied for observing time. PS, GH, DJ, JC, FvW, EP, KB, LP, LSA, and MK observed the targets. DLH extracted single FITS files from data cubes of the SA19+SHOC run. PS did the data reduction for all targets except those observed with DK+DFOSC (EP reduced those data). PS also did photometry and frequency analysis, as well as compiled astrometric parameters of PG~1159 stars and bolometric corrections, derived luminosities, placed the sample in the HR diagrams, and computed radii and pulsation constants. GH and DJ supervised the work. KW provided parts of the data included in Table~\ref{tab:PG1159_physical}. PS wrote the text with contribution from GH and feedback from co-authors. 

\vspace{0.5cm}
We thank Philip Short, Nicholas Humphries, and Martha Tabor who contributed to the observations.
This research was supported in part by the National Science Foundation under Grant No. NSF PHY-1748958 and by the Polish National Center for Science (NCN) through grants 2015/18/A/ST9/00578 and 2021/43/B/ST9/02972. MK acknowledges the support from ESA-PRODEX PEA4000127913.
This paper uses observations made at the South African Astronomical Observatory (SAAO). Based on observations made with the Gran Telescopio Canarias (GTC), installed in the Spanish Observatorio del Roque de los Muchachos of the Instituto de Astrof\'isica de Canarias, in the island of La Palma. Based on observations with the Isaac Newton Telescope operated
by the Isaac Newton Group at the Observatorio del Roque de los
Muchachos of the Instituto de Astrofisica de Canarias on the island
of La Palma, Spain. This paper includes data taken at The McDonald Observatory of The University of Texas at Austin. Data were obtained (in part) using the 1.3 m McGraw-Hill Telescope of the MDM Observatory.
This work has made use of data from the European Space Agency (ESA) mission
{\it Gaia} (\url{https://www.cosmos.esa.int/gaia}), processed by the {\it Gaia} Data Processing and Analysis Consortium (DPAC, \url{https://www.cosmos.esa.int/web/gaia/dpac/consortium}). Funding for the DPAC has been provided by national institutions, in particular the institutions participating in the {\it Gaia} Multilateral Agreement.

\vspace{5mm}
\facilities{GTC (OSIRIS), SAAO: Radcliffe, Elizabeth (SHOC), ING:Newton (WFC), Struve (ProEM), Danish 1.54m Telescope (DFOSC), McGraw-Hill (Andor)}

\software{\texttt{Astropy} \citep{2013AA...558A..33A,2018AJ....156..123A,2022ApJ...935..167A},
          \texttt{ccdproc} \citep{matt_craig_2017_1069648},
          \texttt{dustmaps} \citep{2018JOSS....3..695M},
          \texttt{matplotlib} \citep{Hunter:2007}, 
          \texttt{numpy} \citep{2020Natur.585..357H},
          \texttt{pandas} \citep{mckinney-proc-scipy-2010, reback2020pandas},
          \texttt{Period04} \citep{2005CoAst.146...53L}, 
          \texttt{scipy} \citep{2020SciPy-NMeth}
          }

\bibliography{PG1159survey}{}
\bibliographystyle{aasjournal}

\end{document}